\pdfoutput=1
\documentclass[journal]{IEEEtran}
\usepackage{graphicx}
\usepackage{amsmath}
\usepackage{hyperref}
\usepackage{color}

\DeclareMathOperator*{\argmin}{argmin}

\ifCLASSOPTIONcompsoc
  \usepackage[nocompress]{cite}
\else
  \usepackage{cite}
\fi
\ifCLASSINFOpdf
\else
\fi
\ifodd 1
\newcommand{\rev}[1]{{\color{blue}#1}} 

\newcommand{\torev}[1]{{\color{red} #1}}
\else
\newcommand{\rev}[1]{}
\newcommand{\torev}[1]{}
\fi

\ifodd 1
\newcommand{\com}[1]{\textbf{\color{red} (COMMENT: #1)}} 
\newcommand{\comg}[1]{\textbf{\color{green} (COMMENT: #1)}}
\newcommand{\response}[1]{\textbf{\color{magenta} (RESPONSE: #1)}} 
\else
\newcommand{\com}[1]{}
\newcommand{\comg}[1]{}
\newcommand{\response}[1]{}
\fi

\begin{document}
%
\title{Reliable Physical-layer Network Coding Supporting Real Applications}
%
%
%
%

\author{Lizhao~You,~\IEEEmembership{Student Member,~IEEE,}
        Soung~Chang~Liew,~\IEEEmembership{Fellow,~IEEE,}
        and~Lu~Lu,~\IEEEmembership{Member,~IEEE}
}

\IEEEtitleabstractindextext{%
\begin{abstract}
This paper presents the first reliable physical-layer network coding (PNC) system that supports real TCP/IP applications for the two-way relay network (TWRN). Theoretically, PNC could boost the throughput of TWRN by a factor of 2 compared with traditional scheduling (TS) in the high signal-to-noise (SNR) regime. Although there have been many theoretical studies on PNC performance, there have been relatively few experimental and implementation efforts. Our earlier PNC prototype, built in 2012, was an offline system that processed signals offline. For a system that supports real applications, signals must be processed online in real-time. Our real-time reliable PNC prototype, referred to as RPNC, solves a number of key challenges to enable the support of real TCP/IP applications. The enabling components include: 1) a time-slotted system that achieves  $\mu s$-level synchronization for the PNC system; 2) reduction of PNC signal processing complexity to meet real-time constraints; 3) an ARQ design tailored for PNC to ensure reliable packet delivery; 4) an interface to the application layer. We took on the challenge to implement all the above with general-purpose processors in PC through an SDR platform rather than ASIC or FPGA. With all these components, we have successfully demonstrated image exchange with TCP and two-party video conferencing with UDP over RPNC. Experimental results show that the achieved throughput approaches the PHY-layer data rate at high SNR, demonstrating the high efficiency of the RPNC system.
\end{abstract}

}

\maketitle

\IEEEdisplaynontitleabstractindextext

%
\IEEEpeerreviewmaketitle

\section{Introduction}\label{sec:introduction}


%
%
%
%

 

\begin{figure}
	\centering
	\includegraphics[width=0.5\textwidth]{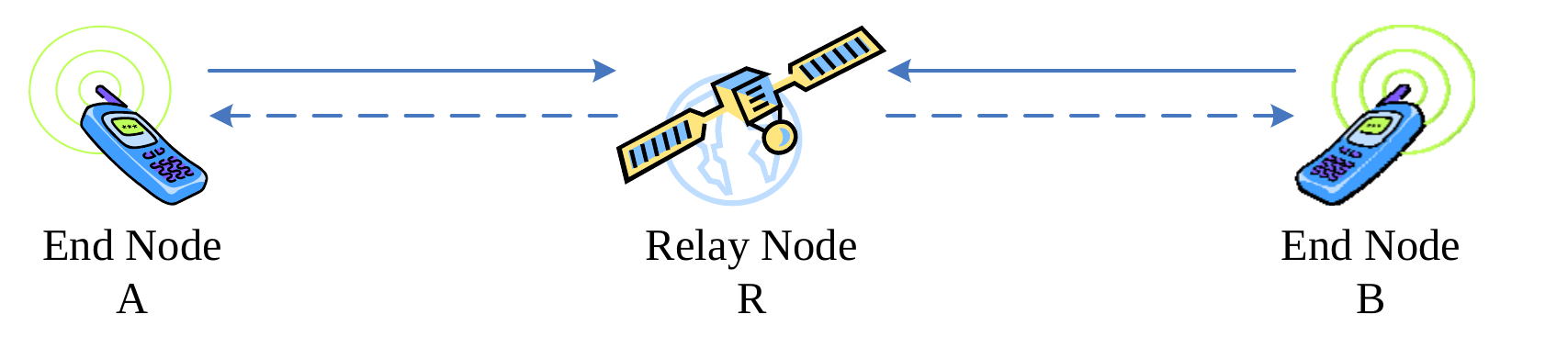}
	\caption{Two-way relay wireless network.}\label{fig:SystemModel}
\end{figure}

Since the introduction of the concept of physical-layer network coding (PNC) \cite{PNC06, PopICC06} in 2006, it has developed into a subfield of network coding with a wide following \cite{PNCSurvey12, NazerSurvey11}. Fig. \ref{fig:SystemModel} illustrates the application of PNC in a two-way relay network (TWRN). In TWRN, two end nodes A and B exchange information via a relay R.  The information exchange consists of two phases (time slots). In the first time slot, nodes A and B send signals simultaneously to relay R. Relay R then processes the superimposed signals of the simultaneous packets and maps them to a network-coded packet (XOR packet). The network-coded packet is then broadcasted to nodes A and B. Each end node then extracts the packet from the other end node by subtracting its own packet from the network-coded packet. Theoretically, PNC could boost the throughput of TWRN by a factor of 2 compared with traditional scheduling (TS) in the high signal-to-noise (SNR) regime \cite{PNC06}. Although there have been many theoretical studies on PNC performance \cite{PNCSurvey12, NazerSurvey11}, there have been relatively few experimental and implementation efforts.

This paper presents the design, implementation and evaluation of the first reliable PNC supporting real applications at the application layer (e.g., file transfer, video conferencing). Our prototype is built on a compute-bound software-defined radio (SDR) platform that performs signal processing using General Purpose Processors (GPP) rather than ASIC or FPGA. 

Four essential elements are critical to the support of real applications on the SDR platform: 1) real-time GPP signal processing (as opposed to offline Matlab-based signal processing); 2) distributed synchronization of nodes running on different clocks/oscillators; 3) an ARQ mechanism to give reliable communication service to the application layer; 4) an interface to the application layer. This paper addresses various challenges arising from implementing and integrating these elements. 

Based on our solutions to these challenges, we built a real-time reliable PNC prototype that can support standard TCP/IP applications, referred to as RPNC. This is an advance from our previous offline-signal-processing prototype in \cite{FPNCPhycom12}.

We make use of the USRP/GNU Radio SDR platform \cite{Ettus}. We focus on USRP/GNU Radio because of its programmability and flexibility for rapid prototyping \cite{IEEEMagazine16}. In this platform, the USRP is the RF frontend hardware. Its receiver samples RF signals and converts them to digital baseband signals. These digital baseband signals are passed to a commodity personal computer (PC), where the GNU Radio resides, via a Gigabit Ethernet. The GNU Radio is the baseband signal processing software. The GNU Radio and other software in the PC also prepare the baseband signals to be transmitted. These digital baseband signals are passed to the USRP, where they are converted to RF signals for transmission by the USRP transmitter. 

In the following, we give a quick overview of the challenges and our solutions to implement RPNC on USRP/GNU Radio.

\subsection{Challenges} \label{sec:intro:challenge}
(a) \emph{Building a Real-Time and High-Throughput System on USRP/GNU Radio}: For data flow between the GNU Radio and USRP, the GNU Radio makes use of a pipelined architecture that glues several signal processing blocks (e.g., FFT, Viterbi decoding blocks) together, the output of one block being the input of the next block. A separate thread is dedicated to the processing of data samples entering each block. The thread is also responsible for passing the processed data samples to the next block. The operating system (OS) is counted upon to schedule the executions of the blocks (threads). 

It is quite challenging to develop a real-time system on USRP/GNU Radio. A well-known problem is the large latency. The latency includes the data transfer latency between the USRP and the PC, and the baseband processing latency in the PC. The time to pass signal samples back and forth between USRP and PC is on the order of $ms$ \cite{NychisNSDI09}. The baseband signal processing time on GNU Radio is also on the order of $ms$ \cite{NychisNSDI09}. Compounding the problem is that this latency is not fixed and has large jitters. The jitters are due to the unpredictable OS scheduling on the multi-threaded GNU Radio, which is a user-space program in a non-real-time OS.

Another challenge is to match the processing speed of the PC with the I/O speed of the USRP.  The USRP hardware produces baseband samples to the host PC at a constant rate. For example, the sampling rate can reach 640Mbps for a 20MHz OFDM system. The software in the PC needs to first identify packets from the massive raw samples coming from the USRP, and then performs additional packet processing (e.g., fast Fourier transform, Viterbi decoding). If the overall computation demands overload the CPU, the signal processing block that copies samples from USRP will drop samples, leading to degraded performance and unreliability.

(b) \emph{Distributed Implementation of Simultaneous Transmissions to within-CP Accuracy}: Our PNC prototype is an OFDM system. In OFDM systems, the transmitter prepends a cyclic prefix (CP) to each OFDM symbol to counter the delay spread caused by multipath channels. The receiver chooses a suitable position (namely a CP-cut) inside the CP as the starting point of the OFDM symbol so that the OFDM symbol is free of inter-symbol interference from the previous OFDM symbol. A proper CP-cut could guarantee that the extracted signals preserve the orthogonality of OFDM subcarriers. 

Our PNC prototype in \cite{FPNCPhycom12} extended the use of the CP. In particular, with respect to the TWRN setup in Fig. \ref{fig:SystemModel}, we showed that if the transmitted signals from nodes A and B to the relay are loosely synchronized, so that the combined delay spread of the signals of the two nodes is within the length of a CP \footnote{Suppose the delay spread of node A is $[\tau_{A,1},\tau_{A,2}]$ (i.e., the first path of the multipath channel for A has delay $\tau_{A,1}$; the last path of the multipath channel has delay $\tau_{A,2}$, and the delay spread of node B is $[\tau_{B,1},\tau_{B,2}]$. The combined delay spread is defined as the interval $[\min(\tau_{A,1},\tau_{B,1}),\max(\tau_{A,2},\tau_{B,2})]$.}, then the asynchrony in their signal arrival times at the relay can be dealt with by the OFDM system in a similar fashion as with multipath fading \cite{JointEstDecod14}. However, if the combined delay spread exceeds the CP length, there will be no ideal CP-cut position to preserve subcarrier orthogonality for both nodes. 

In our previous offline prototype \cite{FPNCPhycom12}, for experimental convenience, we connected two USRPs to the same clock/oscillator. Driven by the same clock, the two USRPs are synchronized in time. We manually set a transmission time, and force these two USRPs to transmit at this time, so that the relative offset between the signal arrival times of the two user nodes is within the CP of OFDM. In a real communication system, the two nodes in the TWRN are not co-located, and the two nodes are driven by two asynchronous clocks. A challenge is to design a distributed protocol that ensures simultaneous transmissions to within-CP accuracy despite the clock asynchrony.

Tackling this challenge is not trivial because of the large latencies and jitters (on the order of $ms$) between the USRP and the GNU Radio. In wideband OFDM systems (e.g., 802.11), the bandwidth is up to tens of MHz bandwidth, and the CP length is usually tens of samples. Therefore, to realize simultaneous transmissions to within-CP accuracy, we need to ensure sample-level synchronization accuracy (i.e., on the order of $\mu s$), far exceeding the $ms$-level timing accuracy on GPP-based SDRs. Furthermore, the jitters prevent precise control of the timing of packet transmissions, making it difficult to realize $\mu s$-level synchronization required by PNC. In short, the PC cannot just prepare the transmit samples and ask the USRP to transmit those samples as soon as the USRP receives them. If the PCs of the two nodes did this, simultaneous transmissions to within-CP accuracy would surely fail due to the unpredictable and different latencies in the two nodes. 

(c) \emph{Building A Reliable and High-Throughput System on USRP/GNU Radio}: To realize reliable communication with an ARQ mechanism at the link layer, the transmitters rely on feedback from the receiver together with timeouts to decide whether to retransmit packets. In commercial wireless systems (e.g., 802.11), the packet processing latency at the link layer is bounded by a pre-defined time duration, and that duration is usually smaller than the packet duration. The transmitters, typically realized in hardware, can leverage the deterministic timing parameters to design ARQ protocols with almost instant feedback.

However, the latency in USRP/GNU Radio is much larger than the packet duration, rendering ARQ that requires instant feedback impossible. Furthermore, the proper setting of timeouts at the transmitter requires a good estimation of the round-trip time, a problem compounded by the unpredictable jitters in USRP/GNU Radio. What is needed is a link-layer protocol for RPNC that could achieve high throughput despite the large latency and jitters.


\subsection{Our Solutions and Contributions}
In the following, we overview our solutions to the challenges. To ease exposition, we begin with challenge (b). 

To tackle challenge (b) of \emph{distributed implementation of simultaneous transmissions to within-CP accuracy}, a natural method is to use a distributed clock synchronization protocol. If such a clock synchronization protocol can achieve $\mu s$-level accuracy, the within-CP requirement can also be met. Existing distributed clock synchronization protocols are either implemented on an application-specific integrated circuit (ASIC) \cite{TDMATMC12} or a field programmable gate array (FPGA) \cite{SourceSync10}. These solutions are not compatible with GPP-based SDRs, where most operations are performed in the PC, whose schedule does not follow deterministic timing sequences as in ASIC and FPGA platforms. 

To realize the tight synchronization, we design a time-slotted system on the USRP/GNU Radio. In particular, we use a distributed clock synchronization protocol that aligns the time-slot boundaries of all nodes in the network to $\mu s$-level accuracy in PC. In our design, the relay's USRP  transmitter clock (i.e., the hardware clock on the relay's USRP board, not the clock in the relay's PC) serves as the reference clock for the overall network. By orchestrating the execution of various processes in the nodes in the network to the time provided by this reference clock, we can realize a distributed protocol. The timing information is conveyed by downlink packets transmitted by the relay. In our slotted system, the arrival times of the downlink packets define the boundaries of the slots. By aligning the slot boundaries at all nodes to the arrival times of the downlink packets, synchronization is achieved. All nodes can only transmit at the beginning of a time slot: thus, if two nodes transmit in the same time slot, their transmissions are aligned to within-CP accuracy, solving challenge (b). 

For a complete solution, the various functions (e.g., preparation of packet samples for transmission) in the PC of all nodes (including the relay) will also need to be synchronized to the slot boundaries, which are set according to the relay's USRP transmitter clock. We develop a tracking method in the PC to maintain slot boundary alignment despite the large data transfer latency. Experimental results show that the synchronization accuracy of our system is within 0.4us.

To tackle challenge (c) of \emph{building a reliable and high-throughput system}, we design a link-layer protocol that adopts a sliding-window ARQ tolerant of the large latency and jitters. In particular, we design an end-to-end link-layer protocol in which only the two end nodes participate in the ARQ mechanism, and that the relay is oblivious of the ARQ. That is, the relay is not responsible for retransmissions and therefore it does not need to store the uplink packets for retransmission purposes. The main reason for adopting the end-to-end design is for simplicity at this stage of our development. The end-to-end link-layer protocol provides a reliable data link for upper-layer applications, and is essential to guarantee a high throughput for TCP applications. Experimental results indicate that, without the link-layer ARQ, TCP could not work at low SNR. Besides that, the TCP throughput without ARQ drops by 50\% at high SNR.

To tackle challenge (a) of \emph{building a real-time and high-throughput system}, we leverage our solutions to challenges (b) and (c): these solutions are part of the solution to challenge (a). More importantly, it turns out the time-slotted architecture can be leveraged to accelerate the signal processing at the PHY layer. In particular, the pre-known packet arrival positions (i.e., at slot boundaries) remove the necessity for searching for potential incoming packets over all time. Experimental results show the CPU utilization is reduced by up to 20\% using the new frame synchronization algorithm in time-slotted architecture. This design makes it feasible for RPNC to support higher data rate for real TCP/IP applications.

Overall, our contributions are as follows:

\begin{enumerate}
\item To the best of our knowledge, we are the first to demonstrate a time-slotted architecture on GPP-based SDR platforms that achieves the 0.4us synchronization accuracy and present a general implementation framework that orchestrates PC operations to the defined slot boundaries despite of the large latency and jitter \footnote{We emphasize that tighter synchronization can be achieved with ASIC and FPGA, but we believe that ours is the state-of-the-art  for GPP-based time-slotted systems.}.  Although our time-slotted system is designed for PNC, we believe the demonstrated time-slotted architecture could also benefit the research of other OFDM-based multi-user systems that leverage simultaneous transmissions to boost throughput, e.g., distributed MIMO and OFDMA.

\item We design and demonstrate a reliable link-layer protocol that incorporates an end-to-end sliding-window ARQ to handle the large latency and jitter on compute-bound GPP-based SDR platforms. To the best of our knowledge, this is the first complete work that addresses link-layer reliability and computation overloading prevention in real-time PNC systems. We believe that the demonstrated protocol could also benefit the implementation of other real-time reliable systems on compute-bound SDR platforms.

\item We demonstrate the first real-time reliable PNC prototype that supports real applications through two examples: the exchange of two image files with TCP and two-party video conferencing with UDP \cite{RPNCRef}. Experimental results show that the achieved throughput approaches the PHY-layer data rate at high SNR, demonstrating the high efficiency of the RPNC system.
\end{enumerate}

Although most of the approaches are related to GPP-based SDR platforms, we believe implementing RPNC on other SDR platforms could also benefit from some of the techniques put forth in this paper. To facilitate system research on PNC and other relevant systems, we release the RPNC source codes, which can be found at \cite{RPNCRef}.

The remainder of this paper is organized as follows: Section \ref{sec:overview} introduces the overall architecture of RPNC. Section \ref{sec:phy} presents the PHY-layer designs, and Section \ref{sec:dll} presents the link-layer designs. Experimental results are given in Section \ref{sec:exp}. Section \ref{sec:related} summarizes related works. Section \ref{sec:conclusion} concludes this paper.

\section{Overview} \label{sec:overview}

This section provides an overview of the RPNC system architecture to give the reader an overall picture on how our RPNC design supports real application in TWRN.

\begin{figure}
	\centering
	\includegraphics[width=0.5\textwidth]{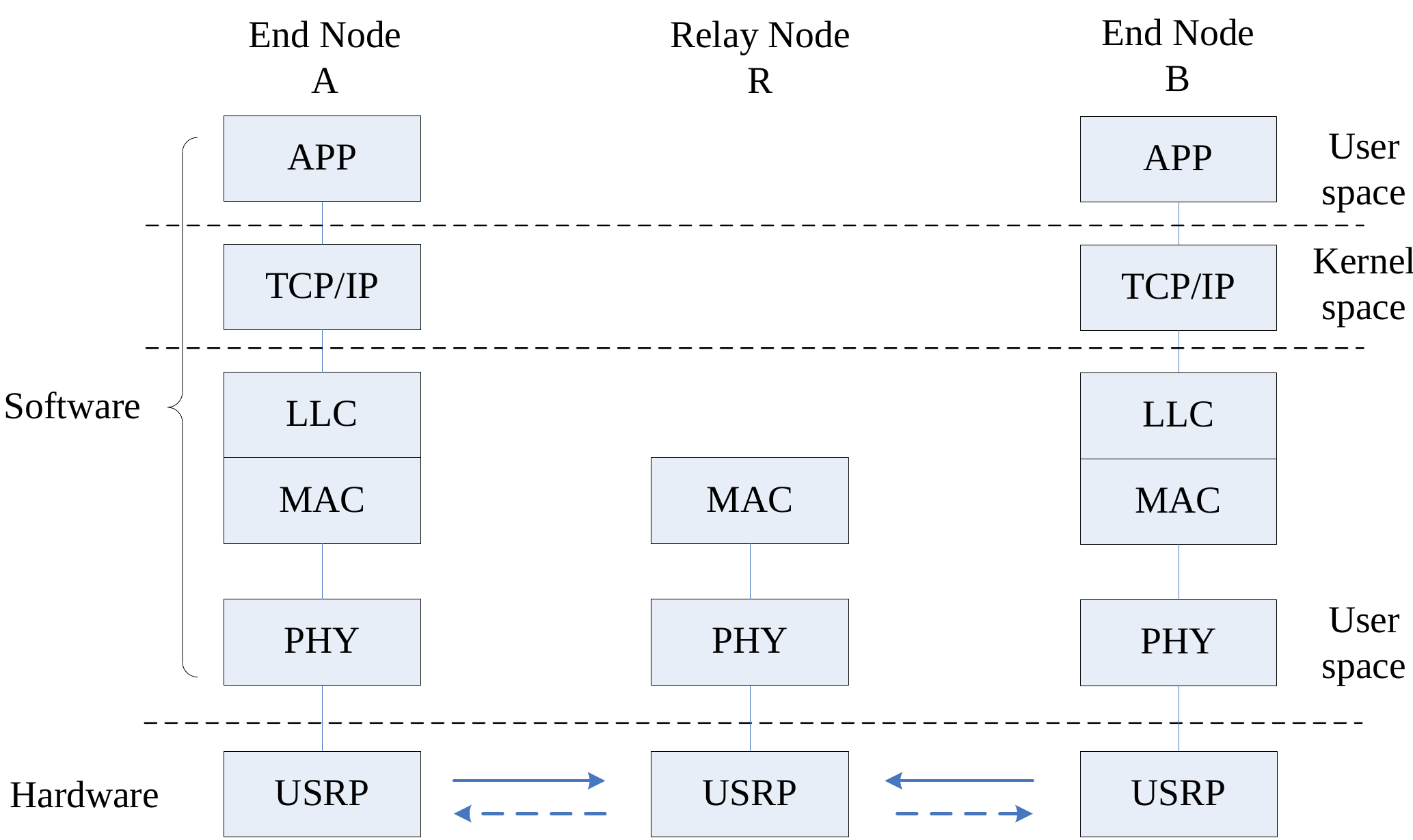}
	\caption{The RPNC system architecture.}\label{fig:Arch}
\end{figure}

Fig. \ref{fig:Arch} shows the architecture of the two end nodes and the relay in the RPNC system. In RPNC, the end node runs real applications over the TCP/IP layer, and the TCP/IP layer interacts with a logic link control (LLC) layer, a medium access control (MAC) layer and a physical (PHY) layer. The relay does not interact with applications directly, and thus it is only equipped with a MAC layer and a PHY layer.

In RPNC, we use the TAP interface \cite{TUNTAP} to simulate a virtual network device. Applications interact with the virtual network device in the same fashion as with other network devices. The virtual network device is implemented in both kernel space and user space. The packets from applications are first delivered to the standard TCP/IP layer in OS kernel. Then the generated Ethernet packets are delivered to a user-space program which attaches itself to the device. We use GNU Radio as the user-space program to deliver Ethernet packets. For the receive path, GNU Radio passes packets to the TAP device, and the TAP device delivers these packets to the TCP/IP network stack and then to applications. 

In addition, the LLC and MAC layers of RPNC are all implemented in GNU Radio. RPNC adopts a time-slotted MAC protocol, where end nodes transmit at the beginning of a time slot, and the slot boundary is maintained by beacon frames sent from relay node. The details will be introduced in Sec. \ref{sec:phy}. To tame the large latency and jitters, RPNC uses a sliding window ARQ for error control at the LLC layer over the time-slotted MAC protocol. The details will be introduced in Sec. \ref{sec:MAC} and \ref{sec:LLC}, respectively. RPNC implements OFDM PHY layer in GNU Radio, and uses the USRP hardware as the RF frontend.

\section{PHY-layer Designs} \label{sec:phy}

This section presents our RPNC PHY-layer designs. Our RPNC system is a time-slotted system in which the two end nodes can transmit to the relay only at pre-defined slot boundaries. The slot boundaries between the two end nodes must be aligned to within certain accuracy for PNC purposes. In particular, since our system is an OFDM system, we need to ensure that the boundaries are aligned to within CP (cyclic prefix) of the OFDM symbols. 

Sec. \ref{sec:phy:slotsync:rx} discusses how the relay and end nodes align their slot boundaries despite the large processing latency incurred by their host PCs. Sec. \ref{sec:phy:slotsync:tx} explains how the relay and end nodes ensure transmissions at established slot boundaries despite large host-PC processing latency. Sec. \ref{sec:phy:framesync} shows how the relay and end nodes implement frame decoding   through frame synchronization at the PHY layer.

\subsection{Slot Boundary Alignment} \label{sec:phy:slotsync:rx}

For our implementation of RPNC, the hardware (USRP) clock at the relay serves as the global reference clock that provides timing reference for the relay as well as the end nodes. Thus, the relay defines its slot boundaries using its own hardware clock (i.e., the end nodes sync to the relay; the relay just uses its own hardware clock to define slot boundaries). In particular, the relay's PC could get the relay's USRP start-up time $T_0$, and define the slot boundaries as $S_n=T_0+nT_s$ for $n=1,2,\cdots$, where $T_s$ is the slot duration.



The end nodes adjust their slot boundaries to synchronize with those of the relay. Specifically, the end nodes use the arrival times of \emph{reference packets} transmitted by the relay to do so. For our current implementation of RPNC, all downlink packets transmitted by the relay serve as reference packets. Besides regular data packets, dedicated packets such as beacon frames can also serve as reference packets. The relay transmits beacon frames during the initialization phase to allow the end nodes to synchronize their slot boundaries with the relay. After that, regular data packets from the relay are used as reference packets to maintain slot alignment. When the relay does not have regular data packets to transmit for a duration of time, it will also transmit beacons to assist the realignment of the slot boundaries at the end node. 


Due to different clocks being used at the relay and end nodes, their slot boundaries may drift apart with time. 
Each end node needs to estimate the drift of its slot boundaries with respect to the slot boundaries of the relay.  
Once the drift exceeds some predefined threshold, the end node will adjust its new slot boundaries going forward.  
By aligning the slot boundaries of the two end nodes to those of the relay, the slot boundaries of the two end nodes are also aligned, allowing the PNC reception mechanism to take effect at the relay. 

Although implementation of the above slot boundary alignment mechanism sounds straightforward on ASIC and FPGA, it is quite challenging on GPP-based SDR due to the large processing latency between hardware and host PC. In the following, we explain how we address the challenge.

\subsubsection{Slot Boundary Estimation and Initialization}

For the RX path of end nodes, a naive idea is to use the time at which a reference packet is decoded on the host PC as its arrival time, $T$. The slot boundaries can then be defined as this arrival time plus some fixed offsets. Specifically, the ensuing slot boundaries could be defined as $S_n=T+nT_s,n=1,2,\cdots$. Unfortunately, this naive idea does not work on the USRP/GNU Radio platform due to the unpredictable data-transfer and decoding latencies between the USRP and the PC (on the order of $ms$). In particular, the processing and decoding latencies incurred by the two end nodes can vary widely. Their slot boundaries may be misaligned by $ms$ because of the discrepancy between their decoding times $T$, violating the within-CP requirement because the CP duration is on the order of $\mu s$. In other words, we cannot use the decoding times registered by the host PCs as the arrival times of reference packets.

To circumvent this problem, we adopt an approach that sets the arrival time to the time at which the reference packet is received in the USRP hardware (i.e., the hardware arrival time). A subtlety, however, is that the USRP hardware does not provide this arrival time directly in an explicit manner. To obtain this arrival time in an indirect manner, the host PC in our RPNC design performs \emph{sample counting}, as explained below. 

\textbf{Principle of Sample Counting}:
the USRP provides a hardware start-up time, $T_0$, when the USRP hardware is turned on initially. This is the time at which the USRP generates its first signal sample in the RX path. Let us refer to this sample of the first USRP sample. The resolution of $T_0$ is on the order of \emph{10ns}. For a system with bandwidth $B$, the USRP generates $B$ samples per second as a continuous stream to the host PC subsequent to the first USRP sample, whether it is actually receiving packets or it is just collecting noise samples. A code segment, referred to as the \emph{counting block}, on the host PC looks for the beginning of a reference packet within this sample stream by doing cross-correlation (see Sec. \ref{sec:phy:framesync}). If it identifies a reference packet, it then locates the first sample of the reference packet. By counting the number of samples between the first USRP sample and the first sample of this reference packet, the host PC can then derive the arrival time T of the reference packet. Specifically, let  be the number of samples between the first USRP sample and the first reference-packet sample. Then $T=T_0+N/B$ \footnote{The sample counting method is viable even if the USRP drops samples due to overflow. This is because when the USRP driver detects an overflow, it will provide a new $T_0$ as a new starting sample.  The sample counting from then on will be performed with respect to the new $T_0$ and the new starting sample.}.

\begin{figure}
	\centering
	\includegraphics[width=0.52\textwidth]{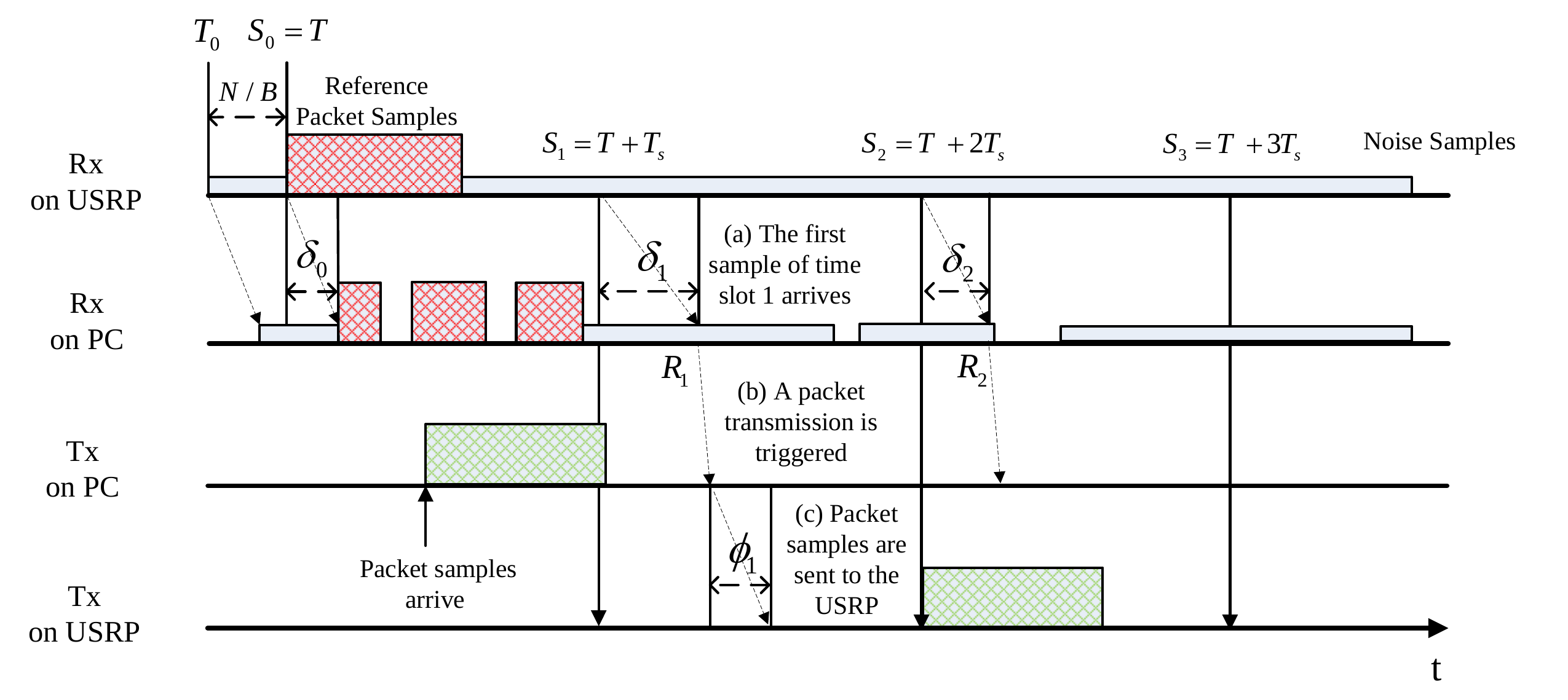}
	\caption{The host-based timing method at an end node facilitated by a reference packet received from the relay. The row labeled by Rx on PC illustrates the process at the counting block on the host PC. The timing gaps on this row correspond to no samples being processed by the counting block at those times.}\label{fig:Timing}
\end{figure}

Each end node uses the very first reference packet it receives to set  the initial slot boundary at $T$. The successive slot boundaries going forward are then $S_n=T+nT_s, n=1,2,\cdots$. The first row of Fig. \ref{fig:Timing} illustrates this timing method for estimating and aligning the initial slot boundaries.

\subsubsection{Slot Boundary Realignment}
Due to different clocks being used at the relay and an end node, their slot boundaries may drift apart with time. In our RPNC design, each end node adjusts its slot boundaries once in a while to realign them with those of the transmitter. Specifically, the arrival times of the reference packets it receives from the relay are used for such realignment. The clocks of the relay and an end node are accurate enough that their slot boundaries will drift apart by more than one sample only after many reference packets. However, there may be estimation errors in detecting the arrival times (the first samples) of individual reference packets, and these errors may cause us to deduce large slot-boundary drift when there is none. To smooth out the estimation errors, instead of adjusting slot boundaries based on the slot-boundary drift estimated from one reference packet, we adjust slot boundaries based on the average of the slot-boundary drifts estimated from a window of reference packets.

Specifically, we estimate the slot-boundary based on a window of reference packets received within a constant number of time slots. Note that the number of reference packets within the window may not be a constant because some time slots in the window may not have reference packets (reference packets may not arrive in a regular manner). This design is motivated by the observation that the clock drift within a time interval is related to the duration of the time interval rather than the number of reference packets received. Also, in our design the end nodes adjust slot boundaries with the granularity of one sample duration (i.e., we do not adjust a slot boundary by a fraction of a sample duration). Each new reference packet does not necessarily lead to adjustment of slot boundaries; only when the aforementioned estimated average slot-boundary drift is at least one sample in magnitude does an end node adjust its slot boundaries. The following specifies our scheme more formally in a quantitative manner. 

The end nodes use non-overlapping windows to calculate the average clock drift. Specifically, the calculations are performed in slots $iW-1,i=1,2,\cdots,\infty$, where $W$ is the window size. The $i$-th window contains slots with indexes in the set $W_i=\{(i-1)W,(i-1)W+1,\cdots,iW-1\}$. Let $I_i \subset W_i$ denote the indexes of slots in which there is a reference packet arrival. The average clock drift of the $i$-th window is computed as
\begin{equation} \label{eqn:a1}
\Delta {T_i} = \left \{ {
	\begin{array}{*{20}{c}}
	{\left\lfloor {\sum\limits_{j \in {I_i}} {({T_j} - {S_j})} /|{I_i}|} \right\rfloor {~~~~~\rm{if}}|{I_i}| > 0}  \\
	{~0~~~~~~~~~~~~~~~~~~~~~~~~~~~~~{\rm{if }}|{I_i}| = 0}  \\
	\end{array},
	} \right.
\end{equation}
where $T_j$ is the arrival time of the reference packet, and $S_j$ is the time at which slot $j$ begins. Here $T_j$ is derived by the sample counting method, and how to implement sample counting will be elaborated in Sec. \ref{sec:phy:framesync}.

Since each calculation may result in a slot boundary adjustment, we add an index $i$ to $S_n$ to represent the new boundaries due to the computed outcome of the $i$-th window, $S_n[i]$. 

When the first reference packet arrives, slot boundaries are initialized to be
\begin{equation} \label{eqn:a2}
S_n[0]=T+nT_s,n=1,2,\cdots,\infty.
\end{equation}

Note that for notational convenience, in the above expression, the tentative time at which slot $n$ begins, for $n=1$ to $n=\infty$, is already set upon the reception of the very first reference packet. The actual times for later slots, however, will be adjusted based on the receptions of subsequent reference packets. Specifically, new slot boundaries after the computation in window $i$ will be adjusted as follows:
\begin{equation} \label{eqn:b}
S_n[i]=S_n[i-1]+\Delta T_i, n=iW,iW+1,\cdots,\infty ~\rm{for}~ i \geq 1.
\end{equation}

Thus, although the tentative slot boundaries are initially set as in Eq. (\ref{eqn:a2}), the actual slot boundaries adopted later must be realigned through Eq. (\ref{eqn:b}) and (\ref{eqn:a1}). As a reminder to the reader, we emphasize that the computation in Eq. (\ref{eqn:a1}) is triggered whenever the sample corresponding to the beginning of time slot $iW-1$ is detected by the host PC. If an adjustment is deemed appropriate, then the RX path adjusts the slot boundaries starting from slot $iW$ based on Eq. (\ref{eqn:b}). Note that slot $iW$ marks the beginning of another window and the new measurement of drift will be based on the new slot boundaries. 

\subsection{Transmission at Slot Boundaries} \label{sec:phy:slotsync:tx}

After the end nodes acquire the slot boundaries of the downlink, the end nodes schedule their uplink transmissions according to these slot boundaries. However, similar to the reception on host PC, a large latency between host PC and USRP prevents us from transmitting instantaneously. That is, to transmit a packet, the host PC needs to schedule a target time slot for the transmission of the packet such that the signal samples will have arrived at the USRP before the expiry of the target time slot.

In our current implementation of RPNC, we do not compensate for the propagation delay induced by the channel and hardware circuit at the end nodes as far as slot alignment is concerned. Therefore, the slot boundaries for the uplink and downlink are only aligned to within-propagation delay accuracy \footnote{When the round-trip propagation delays between the relay and the two end nodes differ widely, then the within-CP requirement for PNC (see Sec. \ref{sec:intro:challenge}) may be violated (because of the offset between the arrival times of the signals from A and B at the relay). In our current implementation (a typical short-range environment similar to that of WLAN), we observe that the difference between the round-trip propagation delays associated with the two end nodes is negligible. Therefore, the end nodes do not need to artificially add extra elastic delays to compensate for the difference between the two propagation delays.}. 

\subsubsection{Ensuring Transmissions at Appropriate Time Slots}
In the USRP setup, the host PC specifies the transmission times of samples that it delivers to the USRP. In other words, the host PC tells the USRP at what times the USRP should transmit these samples. Before the samples of a packet can be forwarded to the USRP, we first need to make sure that the signal samples associated with the packet are ready. We also need to make sure that the samples will arrive at the USRP before their target transmission times; late arriving packets will not be transmitted by the USRP, and will be treated as lost packets in RPNC.

To determine a suitable target transmission time, the host PC faces two challenges: (1) the host PC does not have direct access to the hardware time; (2) there is a data transfer delay from the host PC to the USRP. Therefore, the host PC needs to derive the current hardware time in an indirect manner, and leave a sufficient lag time for the transfer of the samples to the USRP, when specifying the transmission times of samples to the USRP.

In our implementation, the baseband samples of a packet (referred to as \emph{packet samples}) that are ready are put into a transmit queue, \emph{TXQueSample}, at the host PC. Recall that in the RX path, the counting block of the host PC receives RX samples from the USRP in a continuous manner (see Sec. \ref{sec:phy:slotsync:rx}). By sample counting, the counting block knows which samples correspond to the beginnings of slot boundaries. When a sample corresponding to a slot boundary, say of slot $n$,  reaches the counting block, the host PC knows that slot $n$ has transpired in the RX path of the USRP. This triggers a code segment in the TX path to check if packet samples are available in \emph{TXQueSample}. If so, the host PC will specify a time slot later than $n$ for the transmission of the packet samples in the TX path of the USRP.

Let $R_n$ be the current hardware time when the counting block in the RX path identifies the $n$-th slot boundary. The counting block does not know $R_n$ exactly since it has no direct access to the USRP hardware time. Recall that the hardware time of the $n$-th slot boundary is defined as $S_n[i]$, where $i=\lfloor n/W \rfloor$ denotes the window index for slot alignments. The counting block does know $S_n[i]$. We can write 
\begin{equation} \label{eqn:eqn1}
R_n = S_n[i]+\delta_n,
\end{equation}
where $\delta_n$ is the overall delay from the USRP to the counting block in the RX path, including the transfer delay from the USRP to the PC, and the queuing delay in GNU Radio. The host PC should set the transmission time slot of the packet to be $m=\argmin_k \{S_k[i] \geq R_n+\phi_n\}$, where $\phi_n$ is the transfer delay from the PC to the USRP. In general, both $\delta_n$ and $\phi_n$ may vary with $n$. However, if we can bound them using a conservative estimate, say, $\delta>\delta_n$ and $\phi>\phi_n$, then we can set the target transmission time slot to be 
\begin{equation} \label{eqn:eqn2}
m=\argmin_k \{S_k[i] \geq S_n[i]+\delta+\phi \}.
\end{equation}

\subsubsection{Scheduling Transmissions on Host PC - One-Slot Ahead}
The above method requires us to perform measurements to set $\delta$ and $\phi$. In practice, $\delta_n$ and $\phi_n$ may have large jitters \cite{NychisNSDI09}. Also, measuring $\delta_n$ and $\phi_n$ separately is difficult since the clocks on PC and USRP are different. In our implementation, we adopt a method that does not require the measurement of $\delta_n$ and $\phi_n$. In particular, from our experimental data, we find that $\delta_n+\phi_n$ is always smaller than $T_s$, the duration of a time slot, in our implementation. Thus, we simply replace $S_n[i]+\delta+\phi$ in Eq. (\ref{eqn:eqn2}) by $S_{n+1}[i]$. That means when the counting block detects the $n$-th slot boundary, it will schedule the next packet transmission for slot $n+1$. That is, we replace Eq. (\ref{eqn:eqn2}) by
\begin{equation} \label{eqn:eqn3}
m = n+1.
\end{equation}

Fig. \ref{fig:Timing} also shows an example of packet transmissions. At $R_1$, the counting block at the host PC of the end node detects the boundary of slot 1. The host PC then reads packet samples from \emph{TXQueSample}, tags the timestamp $S_2$ to the first sample, and forwards the samples to the USRP; the USRP hardware then buffers these samples, and waits until $S_2$ to transmit the samples (see the green rectangle at the bottom of Fig. \ref{fig:Timing}).  At $R_2$, no packet samples are available for transmission, and thus the end node skips  transmission in time slot 3.

\subsection{Reception at Slot Boundaries} \label{sec:phy:framesync}

As discussed in Sec. \ref{sec:phy:slotsync:rx}, to derive the arrival time of the reference packet $T_j$ in Eq. (\ref{eqn:a1}), the PHY layer of the end nodes must identify the beginning of this reference packet. 
The identification of the beginning of a frame/packet (and hence the signal samples associated with an overall frame/packet) is referred to as frame synchronization. 
For the relay, the frame synchronization is also used for user identification. 
This subsection is devoted to explaining how the relay and end nodes perform frame synchronization to support these functions in time-slotted RPNC systems. In the following, we briefly overview three frame synchronization methods. Their detailed descriptions and complexity analyses  can be found in Appendix \ref{appendix:sync}.


For the offline PNC implementation in \cite{FPNCPhycom12}, cross correlation was adopted. In particular, the two end nodes used identical short training symbols (STS) placed at the same position of the frame format. Cross correlation with respect to the unique sequence in STS was applied on all received samples to locate the starting positions. We refer to this processing mechanism as the \emph{exhaustive cross-correlation} algorithm. The description exhaustive refers to the fact that the cross correlation algorithm must be run over all samples.
	
For real-time PNC systems, a straightforward modification to \cite{FPNCPhycom12} is for the two end nodes to use non-overlapping STS, and to use autocorrelation to identity a rough beginning of a packet. After a rough beginning is identified, then cross correlation can be used to locate a precise beginning in the neighborhood of the rough beginning. This reduces the amount of cross correlation to be performed, hence the complexity is also reduced since cross correlation is more computationally intensive than autocorrelation. We call this modification the \emph{exhausive auto-correlation narrow cross-correlation} algorithm. Since it is a common method in real systems \cite{Sams01}, henceforth we will simply refer to it as the \emph{standard cross-correlation} algorithm.


The standard cross-correlation method does not utilize an advantage brought forth by the time-slotted RPNC system in this paper. In particular, for the time-slotted RPNC system, receivers at the end nodes and the relay can eliminate the autocorrelation computation of incoming signals, because the slot boundaries already indicate the rough starting positions of packets. Whenever the slot boundaries are adjusted, the starting positions are also adjusted accordingly. We refer to the new algorithm as the \emph{narrow cross-correlation} algorithm. This algorithm is used in our current RPNC prototype. In the following, we explain how the end nodes and the relay applies the narrow cross-correlation algorithm. 

For the end nodes, the receivers first perform cross correlation on signal samples of length $C$ in the vicinity of the computed starting position of downlink packets. The role of cross correlation is not only to re-synchronize the end nodes’ clocks to the relay’s clock and to align their slot boundaries, as discussed in Sec. \ref{sec:phy:slotsync:rx}, but also to provide an exact reference point for cutting CP for packet decoding purposes.

The narrow cross-correlation algorithm can be applied only after the slot boundaries are identified. Initially when the system is first started up, an end node does not know where the slot boundaries are. It counts on the transmission of first reference packet (beacon) from the relay to identify the slot boundaries. For this reference packet, the end nodes still use the standard cross-correlation algorithm. After the end nodes receive the reference packet, they switch to the narrow cross-correlation algorithm. 

The standard cross-correlation algorithm is also useful when an end node loses slot synchronization after initialization due to computation errors of Eq. (\ref{eqn:a1}) and (\ref{eqn:b}). In this case, the end node could not identify new reference packets, because the cross correlations are applied at wrong positions. Similar to the initialization process, the end node will switch back to the standard cross-correlation synchronization algorithm to search for a new reference packet whenever it loses slot synchronization. Recall that the relay will always send some reference packets within an interval of time: even when there are no regular data packets to send, it will still send some beacons as the reference packets (see Sec. \ref{sec:phy:slotsync:rx}). Therefore, an end node could detect the loss of slot synchronization if it has not detected any reference packets for a duration time. For example, if the relay is guaranteed to send at least one reference packet every $T_{thresh}$  downlink time slots, an end node could decide that it has lost slot synchronization if it has not received any reference packet in $10T_{thresh}$ (the probability of not being able to detect ten reference packets in succession is very low). 

For the relay, it may appear at first sight that the relay can cut CP based on the uplink slot boundaries, since in time-slotted RPNC, the end nodes align their uplink slot boundaries to the downlink boundaries. A problem prevents us from doing so: there is a small lag time for the uplink slot boundaries as compared with the downlink slot boundaries, because the decoder takes some time to calculate $T_j$ in Eq. (\ref{eqn:a1}). By the time the host PC executes Eq. (\ref{eqn:b}), slot $j$  may have passed. Therefore, the relay also uses cross correlation for frame synchronization in the uplink.

The cross correlation can also be used for user identification. Recall that, if there are no packets in the \emph{TxQueSample} of an end node, it will not send a packet in a time slot. Thus,  for some time slots, only one end node has packet samples to transmit (see Sec. \ref{sec:phy:slotsync:tx}). The relay should be able to detect the absence and presence of transmissions from end nodes, and invoke the appropriate OFDM decoder (i.e., single-user when only one node transmits or PNC when both end nodes transmit). 



In Sec. \ref{sec:exp}, we will demonstrate by experiments that compared with the standard cross-correlation method, the new PHY-layer design for time-slotted RPNC reduces computation and improve system throughput.

\subsection{Transceiver Architecture}  \label{sec:phy:impl}
\begin{figure}
	\centering
	\includegraphics[width=0.4\textwidth]{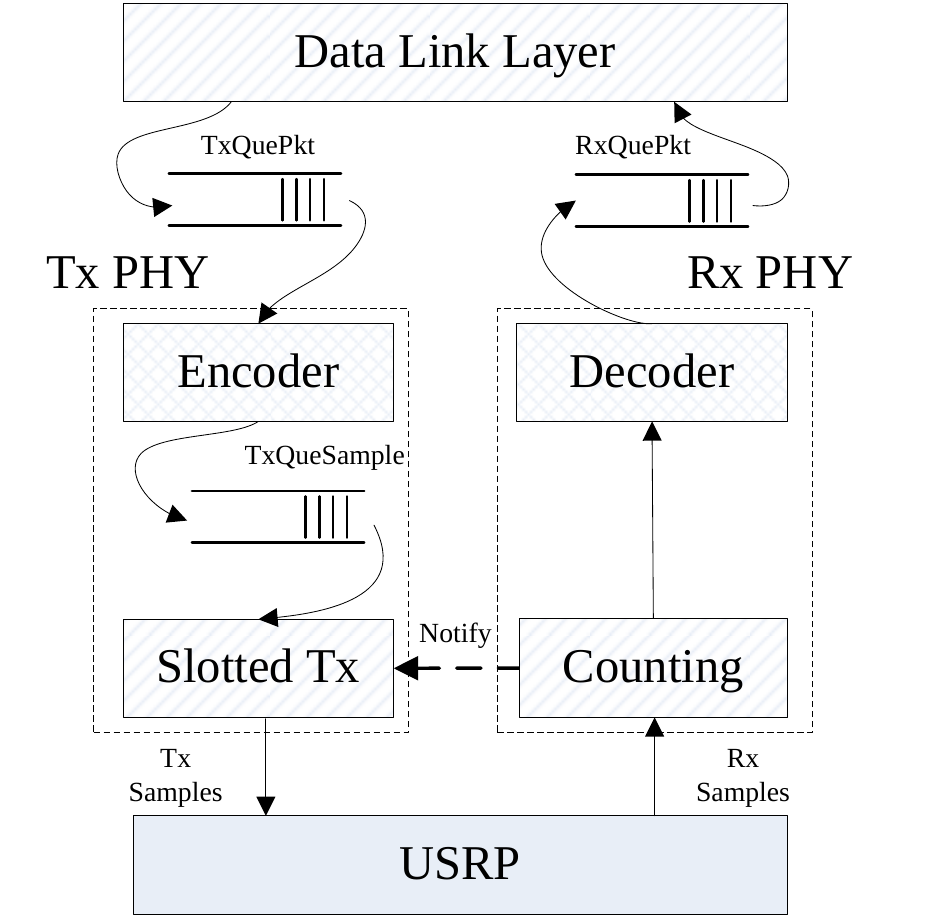}
	\caption{Implementation of an RPNC transceiver. The counting block tracks the slot boundaries, and synchronizes the slotted Tx block to transmit packet samples at appropriate slot boundaries.}\label{fig:PHY-Arch}
\end{figure}

Fig. \ref{fig:PHY-Arch} shows the overall implementation of the RPNC transceiver, focusing on the PHY layers. A transceiver includes three components: the MAC layer, the transmission path (Tx PHY), and the reception path (Rx PHY). The Tx PHY includes the encoder block, and the slotted Tx block. The Rx PHY includes the decoder block, and the counting block. We modify the frame synchronization block in the original OFDM implementation of GNU Radio to serve as the counting block. These blocks are implemented as different threads in the multi-threaded GNU Radio program.
	
There are two queues between the MAC layer and the PHY layer: \emph{TxQuePkt} between MAC and Tx PHY, and \emph{RxQuePkt} between Rx PHY and MAC. A packet from the MAC layer is first put into \emph{TxQuePkt} for PHY-layer encoding; a decoded packet from Rx PHY is put into \emph{RxQuePkt} for MAC-layer processing.
	
The counting block registers $T_0$ by tracking the first signal sample from the USRP. The counting block of an end node relies on frame synchronization to identify the arrival times of received reference packets as specified in Sec. \ref{sec:phy:framesync}, and performs slot boundaries alignment as specified in Sec. \ref{sec:phy:slotsync:rx}.  The counting block of a relay relies on frame synchronization to identify the transmitting end nodes, as specified in Sec. \ref{sec:phy:framesync}.
To support  time-slotted transmissions as specified in Sec. \ref{sec:phy:slotsync:tx}, we introduce a new queue in the stack: \emph{TxQueSample} between the encoder block and the slotted Tx block. The slotted Tx block is synchronized by the counting block. That is, to schedule the transmission in slot $n+1$, the slotted Tx block is first blocked until the counting block approaches  (by the identification of slot $n$ in the RX path). When that happens, the counting block notifies the slotted Tx block. The slotted Tx block reads the available samples associated with a packet, tags the timestamp of $S_{n+1}[i]$, and sends them to the USRP for the transmission in slot $n+1$.
	
\section{Data Link Layer Designs} \label{sec:dll}

To support real applications, PNC systems need a medium access control (MAC) layer to coordinate transmissions of the end nodes and the relay, and a logical link control (LLC) layer to enable reliable data communication. This section presents the details of our RPNC data link layer design.

\subsection{Time-slotted FDD MAC Protocol} \label{sec:MAC}

We propose a time-slotted MAC protocol for data exchange between the two end nodes using the PNC mechanism. The relay transmits downlink packets at downlink slot boundaries. With the end-node time slot synchronization scheme as described in Sec. \ref{sec:phy:slotsync:rx}, the end nodes align their uplink slot boundaries to the downlink slot boundaries. Then the end nodes leverage the scheme in Sec. \ref{sec:phy:slotsync:tx} to ensure the two end nodes transmit only at the beginning of their aligned time slots. In this way, we can ensure simultaneous transmissions to within-CP accuracy for the uplink. 

As described in Sec. \ref{sec:phy:impl}, for the data transmissions, the data packets are first injected into \emph{TxQuePacket} by the data link layer, and then are encoded into packet samples and put into \emph{TxQueSample}. The slotted Tx thread on the host PC checks the availability of packet samples in \emph{TxQueSample}, and transfers the available packet samples to the USRP according to the one-slot ahead rule. In this subsection, our discussions start from \emph{TxQuePacket}. How the data link layer injects the data packets into \emph{TxQuePacket} will be elaborated in Sec. \ref{sec:LLC}.

To simplify MAC design, we use the frequency-division duplex (FDD) mode to isolate the uplink and downlink transmissions. That is, the data from the end nodes are carried on one frequency band, and the data from the relay to the end nodes are carried on another frequency band. 

\begin{figure}
	\centering
	\includegraphics[width=0.5\textwidth]{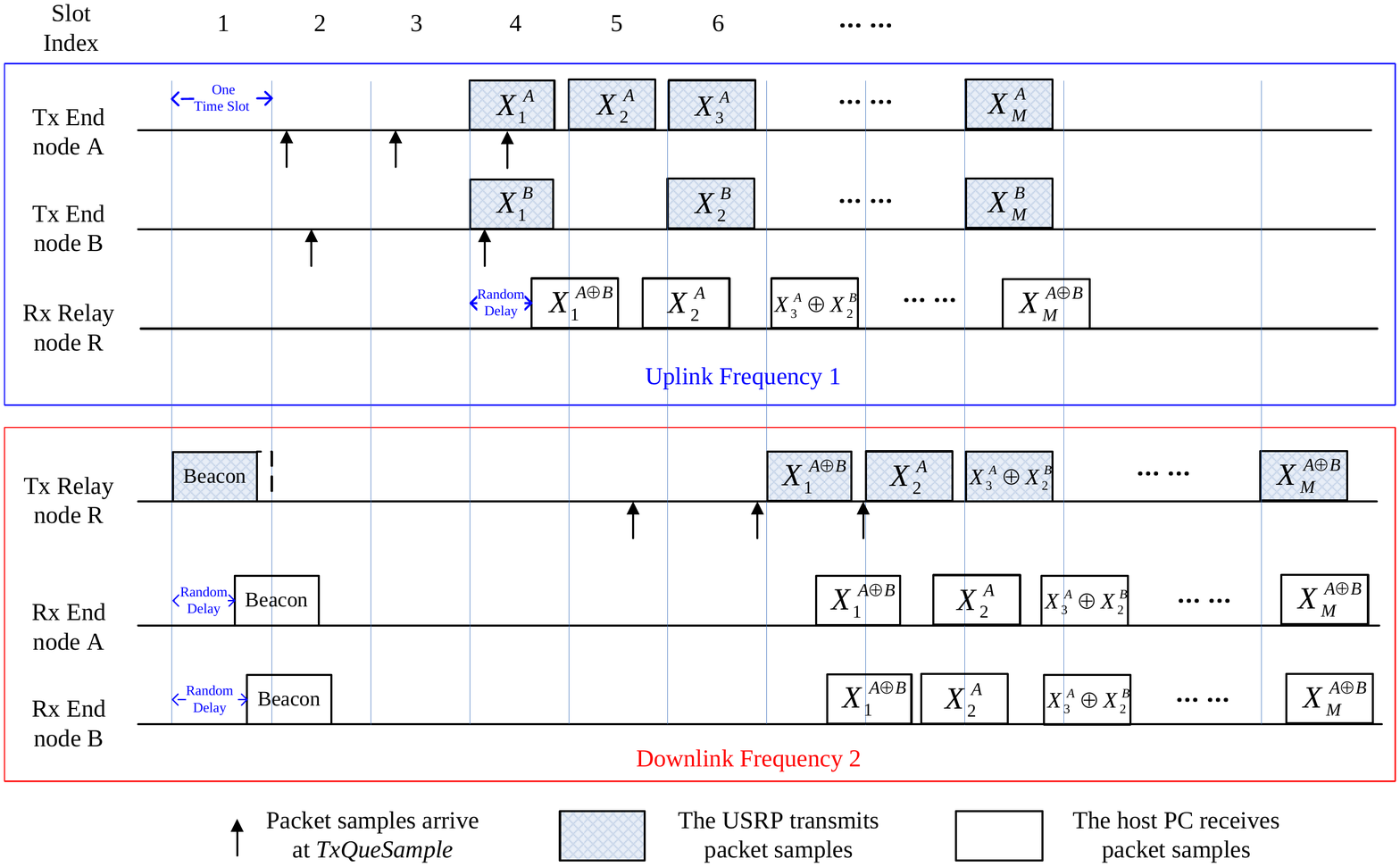}
	\caption{A time-slotted MAC protocol for RPNC.}\label{fig:RPNC-MAC}
\end{figure}

Fig. \ref{fig:RPNC-MAC} shows the operations of the proposed slotted MAC protocol. The uplink and downlink are isolated on two different frequency bands. Let $X^A$ ($X^B$) be the channel-coded packets (i.e., the packet samples) of node A (B) at the PHY layer. The arrows indicate the moments at which the packet samples arrive at the \emph{TxQueSample}. The shaded rectangles indicate the moments at which the packet samples are transmitted by the USRP, and the empty rectangles indicate the moments at which the packet samples are received by the host PC. There exists a non-negligible and random transfer delay between the shaded rectangle and the empty rectangle in USRP/GNU Radio.

To begin, the relay sends a beacon frame, and the end nodes set their slot boundaries according to the beacon frame. Since the end nodes decode the beacon in time slot 2, and set valid slot boundaries from time slot 3, the first uplink transmissions ($X_1^A$ and $X_1^B$) are scheduled for time slot 4 according to the one-slot ahead rule (i.e., the slotted Tx blocks of nodes A and B will check if there is a packet to be transmitted in time slot 4). In the figure, the first downlink XOR packet $X_1^A \oplus X_1^B$ is decoded correctly in the midst of time slot 5 after the slotted Tx block has already waken up and found nothing to be decoded near the beginning of time slot 5. When the slotted Tx block wakes up again near the beginning of time slot 6, it will find $X_1^A \oplus X_1^B$ ready to be transmitted. Thus, the timing block will be schedule its transmission for time slot 7. 

In Fig. \ref{fig:RPNC-MAC}, in time slot 5, node A transmits $X_2^A$, but node B does not transmit because it does not have any packet samples ready to be transmitted (the queue between Tx PHY and the timing layer, \emph{TxQueSample}, is empty). Therefore, relay R only decodes $X_2^A$ correctly (in time slot 6). Since RPNC uses the FDD setup, relay R transmits $X_2^A$ alone without waiting for packets from node B (in time slot 8). In time slot 6, node A transmits $X_3^A$, and node B transmits $X_2^B$. Relay R decodes $X_3^A \oplus X_2^B$ correctly (in time slot 7), and transmits $X_3^A \oplus X_2^B$ (in time slot 9). Transmissions and receptions in other time slots follow the same rules.

Note that there are several idle slots at the beginning of Fig. \ref{fig:RPNC-MAC} (i.e., slots 2 and 3 in the uplink; slots 2 to 6 in the downlink). This is due to actions (synchronization in the uplink and empty \emph{TxQueSample} in the downlink) following the first reference packet. Later on, for subsequent reference packets, RPNC will operate in a pipeline manner, eliminating these idle slots. 

\subsection{ARQ Protocol} \label{sec:LLC}

Automatic Repeat reQuest (ARQ) is a common approach to provide error control at the LLC layer. This subsection is devoted to the discussion of ARQ tailored for RPNC.

\subsubsection{Overview}

\textbf{End-to-End ARQ}: Recall that PNC has two phases: the uplink phase and the downlink phase. The relay first decodes the packets in the uplink, and then forwards the correct uplink packets in the downlink. Therefore, depending on what role the relay plays in the error control, there are two different ARQ models. 

\emph{ARQ-oblivious Relay}: The first model is an end-to-end ARQ model, in which the relay does not participate in the ARQ process. That is, the relay simply forwards any correctly received packets. If the uplink packets are erroneous, the relay does not request retransmissions from the end nodes. In addition, the relay does not store forwarded packets for future retransmissions in the downlink (i.e., it forwards a packet only once). Thus, errors in both uplink and downlink can trigger retransmissions by the end nodes. The receiving end node request retransmissions, and the sending end node retransmits, when there are missing packets. 

\emph{ARQ-participating Relay}: The second model is an link-by-link ARQ model. The relay can request retransmissions if the uplink packets are in error. In addition, the relay can store the transmitted downlink packets, and do retransmissions if any of the end nodes fail to receive the downlink packets. In this case, the uplink and the downlink operate two independent ARQ processes that are decoupled from each other: the loss of the uplink transmissions triggers retransmissions at the end nodes; the loss of the downlink transmissions triggers retransmissions at the relay. 

In this paper, RPNC adopts the end-to-end ARQ for simplicity at this stage of our development. 

\textbf{Sliding Window ARQ}: To provide ARQ error control, the receiver needs to transmit feedback packets after receiving the data packets from the transmitter. Based on the feedback packets, the transmitter decides whether to retransmit the departed packets. Besides feedback packets, retransmissions may also be triggered whenever a timer times out without receiving any feedback packet for a while (note that feedback packets can be lost due to errors as well). 

There are two possible ARQ protocols: the stop-and-wait ARQ and the sliding window ARQ. In the stop-and-wait ARQ, the transmitter is only allowed to transmit one data frame until a feedback frame comes back. While in the sliding window ARQ, the transmitter is allowed to transmit several data frames that are within a window continuously without receiving a feedback. 

The stop-and-wait ARQ is usually used in physical links when the round trip time (RTT), is small compared with  the packet transmission time. For example, in 802.11 wireless LANs, an immediate feedback (either an ACK frame or an idle channel) follows a data frame in the distributed control function (DCF) mode. The sliding window ARQ is usually used in physical links with large RTT, such as Satellite networks, because the sliding window ARQ allows a window of transmissions during the long delay in feedback, and is therefore more efficient. 

In GPP-based SDRs, in addition to the propagation delay, the data transfer, encoding and decoding latencies also contribute to the overall round trip time. Furthermore, since RPNC adopts the end-to-end ARQ, the round trip time is doubled when a data frame or an ACK frame is forwarded by the relay. Therefore, RPNC adopts the sliding window ARQ, given that the RTT is non-negligible compared with the packet transmission time. 

\subsubsection{SR ARQ with SACK} \label{sec:llc:arq}
To provide error control functionality, the data link layer (of the end nodes and the relay) must be able to detect errors on received packets. We adopt the same CRC (cyclic redundancy check) function defined in the IEEE 802.11 \cite{dot11std07}. However, for proper use of the CRC on XOR packets (for uplink in PNC), a subtlety is involved, as explained in Appendix \ref{appendix:crc} \footnote{In general, the CRC function is linear, and the relay could perform CRC directly on XOR packets to detect errors. However, we find that the CRC32 function defined in IEEE 802.11 is non-linear, and therefore we modify the error detection algorithm for relay.}.

RPNC uses a Selective Repeat (SR) ARQ with Selective Acknowledgement (SACK) similar to that in TCP \cite{RFC2018}. In this ARQ protocol, the receiver buffers out-of-order data packets, and returns cumulative acknowledgement. Meanwhile, the receiver also acknowledges discontinuous blocks of out-of-order packets that were received correctly. A block is a group of successive packets that are correctly received.

\textbf{RTT Estimation and Timeout}: The ARQ in PNC needs to estimate RTT to set the  timeout interval and the window size. We adapt the method of TCP for our purposes here \cite{RFC2018}.

Each end node maintains two variables, $RTT_{sample}$ and $RTT_{est}$, for RTT estimation. The sample RTT, denoted by $RTT_{sample}$, is the amount of time between when a tagged packet is sent and when an acknowledgement for the tagged packet is received. Since $RTT_{sample}$ will fluctuate from packet to packet due to the varying PC processing speed, an exponential average RTT, $RTT_{est}$, is introduced to smooth out the fluctuation. Upon obtaining a new $RTT_{sample}$, the transmitter updates $RTT_{est}$ according to the following formula:
$$RTT_{est}=(1-\alpha)RTT_{est}+\alpha RTT_{sample},$$
where $\alpha$ is a weighting factor.

The transmitter also tracks the variability of the RTT, denoted by $RTT_{dev}$, using the following formula:
$$RTT_{dev}=(1-\beta)RTT_{dev}+\beta|RTT_{sample}-RTT_{est}|,$$
where $\beta$ is a weighting factor. 

Then the timeout interval is chosen as $RTT_{est}+4*RTT_{dev}$.

\textbf{Window Size}: Unlike TCP, which is designed for the transport layer, RPNC does not need to consider \emph{congestion control} at the link layer, because the traffic from A always goes to B, and the traffic from B always goes to A, and generally there is no congestion in the two-way relay network within the air channel. Therefore, the window size can be set as  $\lceil (RTT_{est}+4*RTT_{dev})/T_s \rceil$.

\textbf{Flow Control}: In a  compute-bound platform, flow control needs to be considered. The reason is that, if the transmitter sends packets too fast, it may overwhelm the receiver’s processing ability (even though the wireless link may support the traffic), leading to performance degradation due to computation overloading.

However, for implementation simplicity, for the work reported in this paper,  we found a suitable slot duration by experimental adjustment so that no overflow of input samples occurred under saturated traffic, and then developed the link-layer protocol without considering flow control. We will detail  how we set the slot duration through experiments in Sec. \ref{sec:exp}.

\textbf{Piggyback}: Bidirectional traffic is a scenario in which PNC has a substantial advantage over traditional relaying.  In this situation, node A has DATA packets for node B, and node B has DATA packets for node A. Thus, node A needs to transmit ACK to node B, and node B needs to transmit ACK to node A, for DATA traffic in the reverse directions. With bidirectional traffic, we could piggyback the ACK inside the DATA packets in the reverse direction. Accordingly, RPNC adopts a packet format that contains both DATA and ACK fields in the header, and uses flag bits to indicate whether it contains ACK only, DATA only, or both ACK and DATA. The details are elaborated below.

\subsection{Implementation Details}

\begin{figure}
	\centering
	\includegraphics[width=0.5\textwidth]{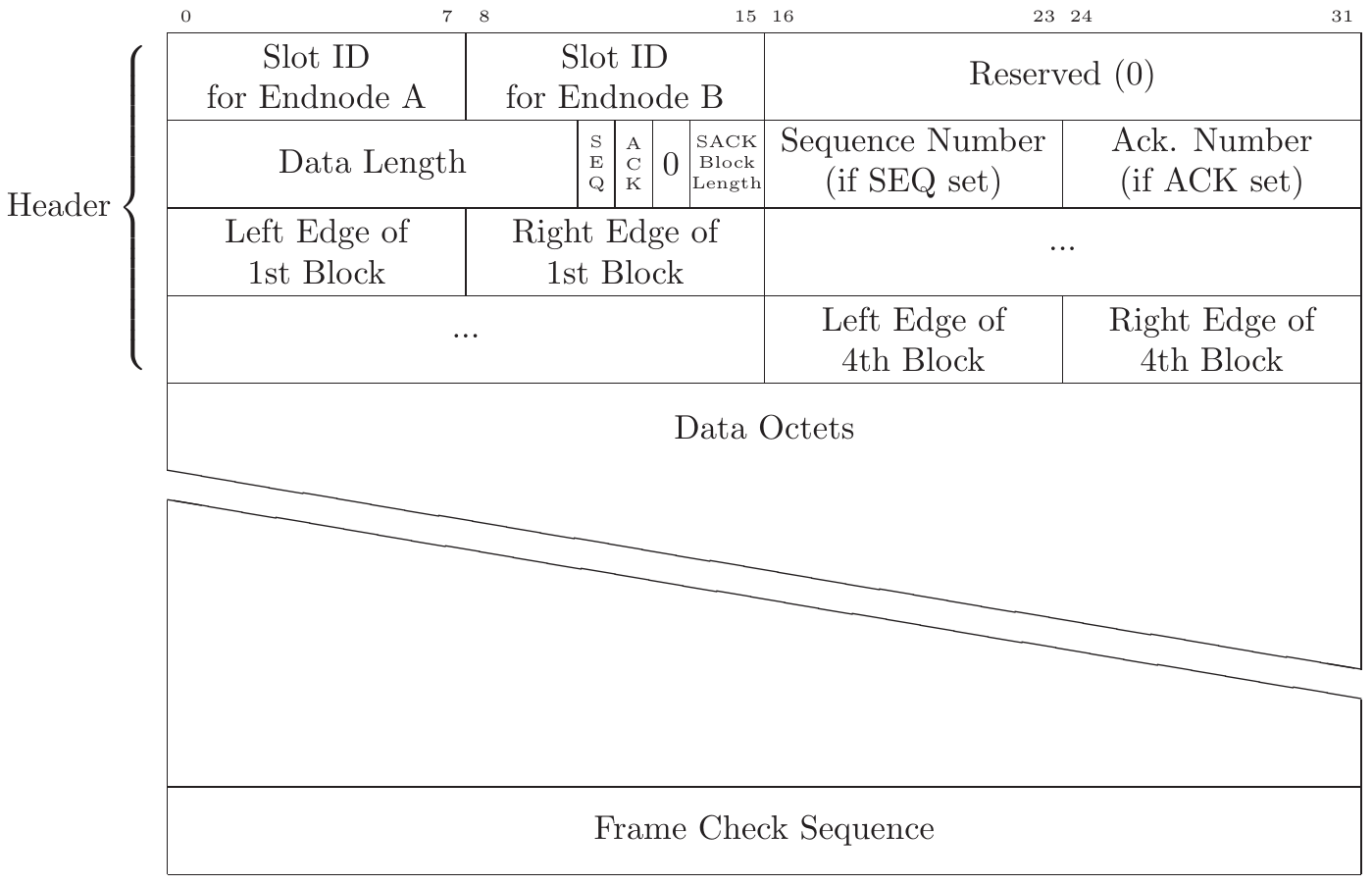}
	\caption{RPNC packet format at the data link layer.}\label{fig:PacketFormat}
\end{figure}

\textbf{Packet Format}: Fig. \ref{fig:PacketFormat} shows the RPNC packet format at the data link layer. The same packet format is used for uplink and downlink. A packet consists of a header, a data section, and a frame check sequence (i.e., CRC).  The size of the header is 16 Bytes, with some fields reserved for future extension. The data section follows the header. The CRC section is 4 Bytes. The valid data length is defined in the data length field in the header.  

Let us first describe the uplink usage of the fields. The only fields that are exclusively used by either node A or node B are the Slot ID fields. The other fields (including the other header fields, the DATA field, and the CRC field) may contain overlapped bits when both nodes A and B transmit simultaneously. Node A will set the first byte with the slot ID when it transmits, and node B will set the second byte to the slot ID when it transmits. There are 11 bits in the data length field, allowing a maximum payload size of 2048 Bytes. We pad dummy bits if the size of data from the higher layer is less than the fixed size. To simplify the CRC calculation of XOR packets at the relay, the position of CRC section is fixed, and the data section size is also fixed. 

The SEQ and ACK flag fields are in the sixth byte. If this packet contains data, the SEQ field is set to 1, and the sequence number is put in the seventh byte. If this packet contains ACK, the ACK field is set to 1, and the acknowledge number is put in the eighth byte. Fields starting from the ninth byte are for SACK, allowing at most four blocks to be acknowledged. The block length is specified in the SACK block length field (after the SEQ/ACK fields) with length 2 bits.

We next describe the downlink. Recall that RPNC uses an end-to-end ARQ. The relay simply forwards any correctly received uplink packets. Therefore, the forwarded packet is either the same as the uplink packet (if only one user transmits in the uplink), or the XOR packet from two uplink packets (if two users transmit simultaneously in the uplink). 

Besides forwarding packets, the relay may need to inject beacons to assist the realignment of the slot boundaries at the end nodes (when there are no packets to be forwarded for a duration of time). We simply reuse the same packet format for the beacons, and send a beacon packet with all bytes in header and data sections set to 0.

\textbf{Downlink Packet Processing}: In RPNC, downlink packet processing is different from the packet processing in traditional point-to-point communication system. In particular, there are three types of downlink packets: XOR packets, single-user packets and beacon packets. An end node needs to identify the packet type, and processes the packet accordingly. 

An end node examines the Slot ID fields to identify downlink packet type and self-transmitted packet. When a downlink packet contains no packet originating from node A (node B), the corresponding Slot ID field is set to zero. Any nonzero slot ID value indicates there is an embedded packet from the corresponding source. Thus, by examining the two Slot ID fields, an end node can find out the packet type. If the Slot ID field associated with the end node is nonzero, it also identifies the self-information. Thus, in case of an XOR packet, the end node can subtract the self-information from the XOR packet to obtain the information coming from the other end node. Specifically, the following operations apply:
\begin{itemize}
	\item If both its slot ID field and the other end node’s slot ID field are not 0, then it is an XOR packet. The end node first obtains its transmitted packet through its own slot ID. Then it performs the XOR operation to extract the packet from the other end node. The extracted packet will be processed at the link layer.
	\item If its slot ID field is 0 but the other end node’s slot ID field is not 0, then the packet is from the other end node. The packet will be processed at the link layer directly.
	\item If its slot ID field is not 0 but the other end node’s slot ID field is 0, then the packet is from itself. The packet will be discarded at the link layer.
	\item If its slot ID field is 0 and the other end node’s slot ID field is 0, then the packet is a beacon. It is used for synchronization purpose only, and does not contain data. The packet will be discarded at the link layer. 
\end{itemize}

\section{Experimental Evaluation} \label{sec:exp}

For the work in this paper, we adopt a PHY-layer implementation similar to that  in \cite{NCMA15}. Here, we focus on tackling system challenges to support real TCP/IP applications. Our current implementation can support real-time operation with 5MHz channel bandwidth. Each packet consists of 512 OFDM data symbols and fields for 4 OFDM training symbols (see Fig. \ref{fig:FrameFormat} in Appendix \ref{appendix:sync}). The data are BPSK modulated, and a rate 1/2 convolutional code is used. Hence, the packet duration is 8.256ms, and the packet size is 1536 Bytes. We choose this number of data symbols to ensure the packet size is larger than the maximum Ethernet frame size outputted by the TAP interface (i.e., no fragmentation in link layer). We use the FDD mode. The uplink uses RF of 2.585GHz, and the downlink uses 2.185GHz.

For our experiments, we deployed three sets of USRP N210 with on-board TCXO clocks and SBX daughterboards \cite{Ettus} in an indoor office environment to emulate a TWRN system. The USRP N210 was connected to a PC (with Intel Core i7-3770 3.4GHz, 16G RAM) through Gigabit Ethernet. We used UHD v003.008.005 to drive the USRP hardware, and GNU Radio v3.6.5 to develop the full protocol stack. The PC ran Ubuntu 12.04.3 LTS with Linux 3.8.0-33-generic kernel.

We first performed experiments to demonstrate reduced computation with the time slotted architecture, varying the time slot durations to find the smallest time-slot duration for which the system did not saturate due to computation limit. After that, we used the smallest time slot thus found for further experiments. Then, we measured the slot synchronization accuracy in our time slotted system. Finally, we obtained the TCP/UDP throughput results on our prototype, and compare the real-time performance with the traditional scheduling system.

\subsection{Reduced Computation in Time Slotted Architecture}

Recall that the time slotted architecture obviates the need for  performing auto correlation all the time, as presented in Sec. \ref{sec:phy:framesync}. The design concept reduces the computation complexity for our compute-bound USRP/GNU Radio platform.

\textbf{Method}: We compare two frame synchronization algorithms: (a) the standard cross-correlation algorithm, a straightforward frame sync implementation for RPNC; (b) the narrow cross-correlation algorithm, the implementation for time slotted RPNC. In addition, we control the duration of time slot $T_s$ to emulate different traffic load. Given a particular $T_s$, we record the decoded PC time of each packet as the arrival time, and measure the arrival gap between successive packets at both end nodes and relay. If the arrival gaps remain constant, then the system could keep up with the processing demand, and is stable. If the arrival gap grows unbounded, then the system is saturated as far as computation is concerned.

\begin{table} \label{table:EndnodeCPU}
	\centering
	\caption{Overall CPU utilization at end nodes.}
	\begin{tabular}{|c|c|c|} \hline
		$T_s$ & Standard cross-correlation &  Narrow cross-correlation \\ \hline
		$10ms$ & 94.99\% & 74.92\% \\ \hline
		$20ms$ & 63.79\% & 47.73\% \\ \hline
		$30ms$ & 50.40\% & 44.29\% \\ \hline
		$50ms$ & 42.13\% & 35.94\% \\ \hline
		$100ms$ & 36.31\% & 31.78\% \\ \hline
	\end{tabular} 
\end{table}

\begin{table} \label{table:RelayCPU}
	\centering
	\caption{Overall CPU utilization at relay.}
	\begin{tabular}{|c|c|c|} \hline
		$T_s$ & Standard cross-correlation &  Narrow cross-correlation  \\ \hline
		$10ms$ & 96.79\% & 82.88\% \\ \hline
		$20ms$ & 75.76\% & 71.45\% \\ \hline
		$30ms$ & 65.55\% & 60.67\% \\ \hline
		$50ms$ & 57.46\% & 51.71\% \\ \hline
		$100ms$ & 55.77\% & 47.23\% \\ \hline
	\end{tabular} 
\end{table}

\textbf{CPU Utilization}: We use the software SYSSTAT \cite{SYSSTAT} to measure the overall CPU utilization. We run the measurement for 300 seconds, and get the overall average results. For implementation simplicity, we vary the slot duration to be an integral multiple of 10ms. 

Table \ref{table:EndnodeCPU} shows the overall CPU utilization of end nodes. The CPU utilization reduction ranges from 5\% to 20\%. Note that the CPU utilization is the sum total of the fractions CPU times consumed by all signal processing blocks and other accessory functions. The correlation process is just one of the many blocks. Thus, an improvement of 5\% to 20\% can be considered as rather significant. When  , the narrow cross correlation algorithm reduces the CPU utilization significantly (i.e., by 20\%). 

Table \ref{table:RelayCPU} shows the overall CPU utilization for the relay. Similarly, the CPU utilization reduction ranges from 5\% to 14\%. The improvement is also significant when $T_s=10ms$. In addition, we observe that the CPU utilization at the relay is larger than that at end nodes given the same $T_s$. This is consistent with our understanding that the decoder for the relay is more complex than that for the end nodes. 

\begin{figure}[tbp]
	\centering
	\includegraphics[width=0.45\textwidth]{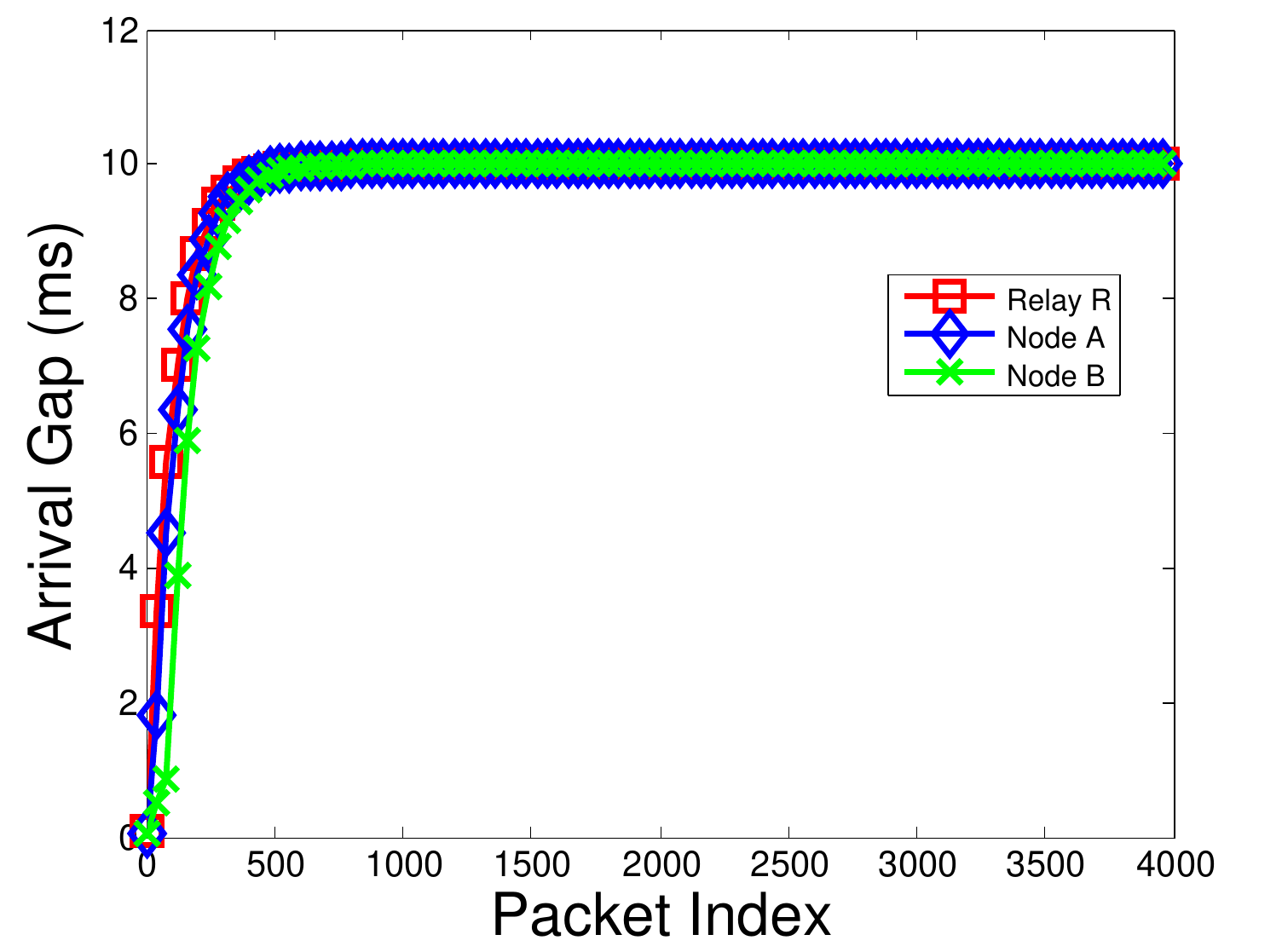}
	\caption{Smoothed arrival gap of decoded packet for $T_s=10ms$.}\label{fig:SystemStability}
\end{figure}

\textbf{System Stability}: Fig. \ref{fig:SystemStability} shows the smoothed arrival gap of decoded packets versus packet index for $T_s=10ms$. The weighting factor is 0.01. As shown in the figure, as time goes, the arrival gaps of successive packets remain constant, and are close to the slot duration -  the packet input gap. Since the arrival gap is based on PC clock and the packet input gap is based on hardware clock, we could not directly compare their value. However, we do not observe overflow during the operation of the program. That is, the computation does not overload the whole system, and the whole system is stable with respect to the input rate $T_s=10ms$. Given the packet duration is at most $T_p=8.256ms$, $T_s=10ms$ is near the optimal performance in PHY layer. 
	
The above results demonstrate the reduced computation complexity of the time-slotted architecture. In the following experiments, we will use the narrow cross correlation algorithm, and set the slot duration to 10ms.

\begin{figure*}[t]
	\begin{minipage}{0.32\linewidth} 
		\centering
		\includegraphics[width=0.99\textwidth]{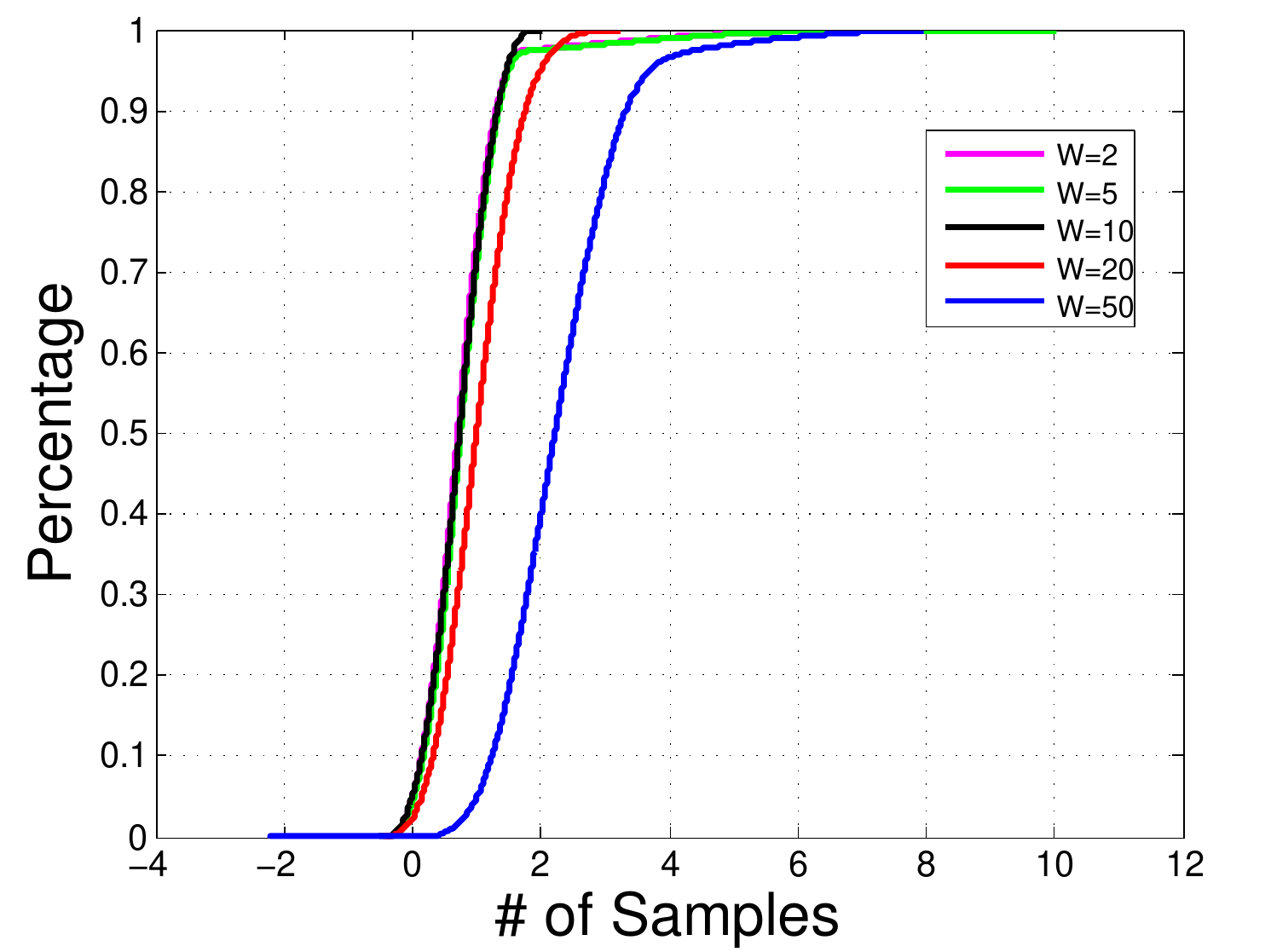}
		\caption{Signal arrival difference of nodes A and B over $W$.}\label{fig:SyncAccuracyRxW}
	\end{minipage}
	\hfill
	\begin{minipage}{0.32\linewidth} 
		\centering
		\includegraphics[width=0.99\textwidth]{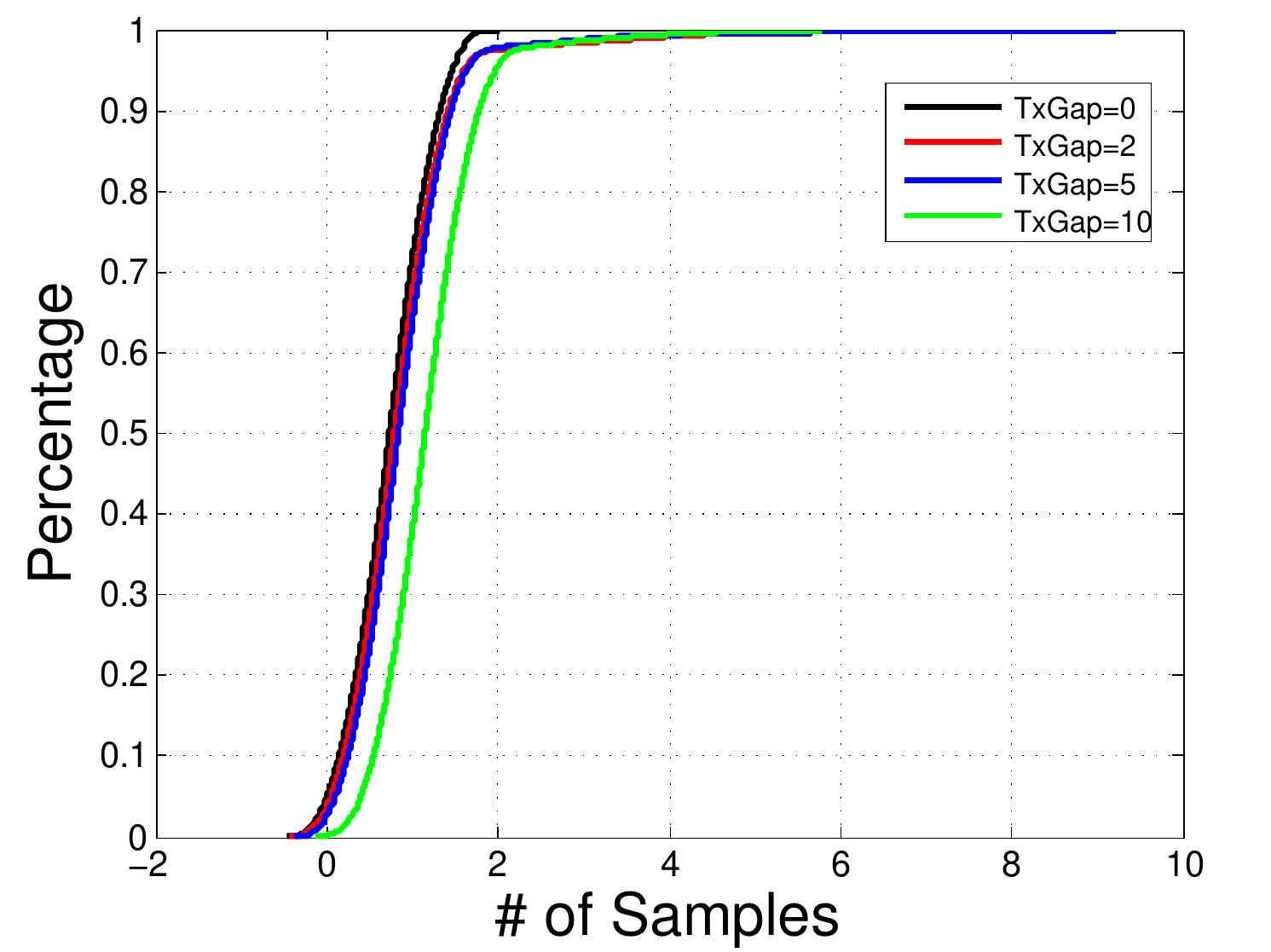}
		\caption{Signal arrival difference of nodes A and B over $TxGap$.}\label{fig:SyncAccuracyTxW}
	\end{minipage}
	\hfill
	\begin{minipage}{0.32\linewidth} 
		\centering
		\includegraphics[width=0.99\textwidth]{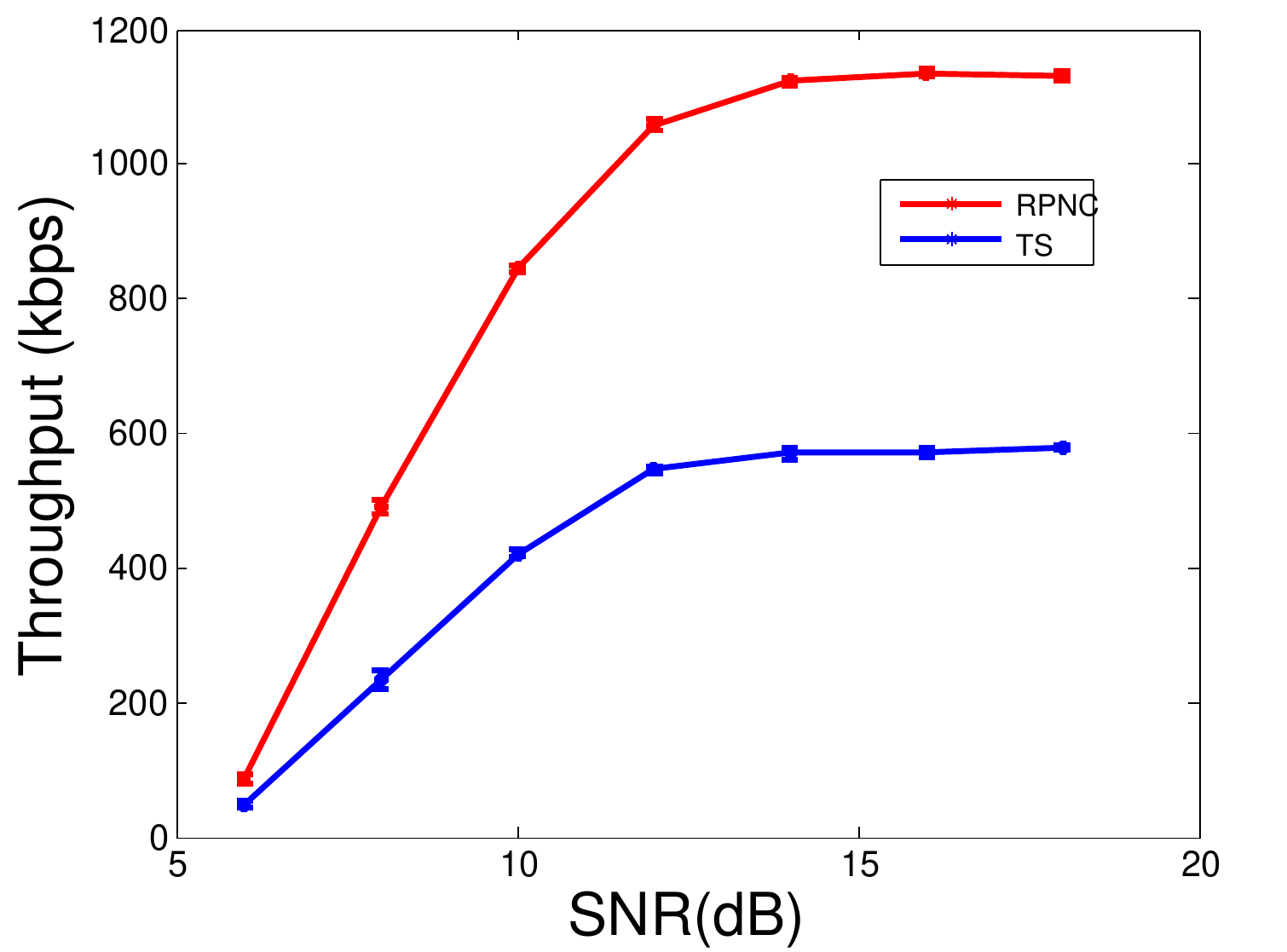}
		\caption{UDP throughputs over different SNRs for RPNC without ARQ and TS without ARQ.}\label{fig:UDPThroughput}
	\end{minipage}
\end{figure*}

%
%

\subsection{Slot Synchronization Accuracy}
\textbf{Method}: In this experiment, the end nodes and the relay use the time-slotted architecture. We set the end nodes to transmit packets in each time slot, and measured the arrival-time difference of the packets of nodes A and B at the relay. 

To measure the difference with sub-sample resolution, we leverage the property of FFT processing in OFDM: a $\Delta$ sample delay introduces a phase shift to an OFDM subcarrier proportional to the subcarrier index, and the slope of the phase shift versus subcarrier index curve is $\zeta=\frac{2\pi\Delta}{N_s}$, where $N_s$ is the number of subcarriers in an OFDM symbol. Recall that the relay can obtain the CSI of both node A (i.e., $CSI_A$) and node B (i.e., $CSI_B$) through orthogonal LTS. We can regress the slopes $\zeta_A$ and $\zeta_B$ (also $\Delta_A$ and $\Delta_B$) from $CSI_A$ and $CSI_B$. Then, the arrival-time difference is given by $\Delta_d = \Delta_A-\Delta_B$.

\textbf{Results}: Fig. \ref{fig:SyncAccuracyRxW} shows the cumulative distribution function of $\Delta_d$ over different $W$. Recall that $W$ is the non-overlapping window used to calculate the average clock drift. If $W$  is too small, then the clock drift calculation may not be accurate. On the other hand, if $W$ is too large, then it will take a long time to trigger the slot boundary adjustment. Both lead to large  $\Delta_d$.  

According to Fig. \ref{fig:SyncAccuracyRxW}, $W=10$ is an appropriate window size in our system, and the arrival-time difference is less than 1.5 samples (i.e., 0.3us) with 90\% probability.

Recall that if the relay does not have sufficient downlink data packets, it needs to send reference packets regularly so that the end nodes could maintain accurate slot boundaries. Define $TxGap$  to be the number of time slots between successive downlink reference packets. In the below experiment, we study the impact of different $TxGap$ on the synchronization accuracy. After that, we use the largest $TxGap$ thus found for protocol design to ensure accurate synchronization. 

Fig. \ref{fig:SyncAccuracyTxW} shows the cumulative distribution function of $\Delta_d$  over different $TxGap$. As shown in this figure, when $TxGap=0$, the system achieves the best synchronization accuracy; when $TxGap$ increases, the signal arrival-time difference also increases. However, even when $TxGap=10$ (i.e., around 100ms, same as the beacon interval in 802.11), the difference is less than 1.8 samples (~0.36us) with probability larger than 90\%. In our system, most of time, downlink XOR packets can serve as synchronization packet. If not, we ensure that there is at least a downlink packet every 10 time slots, by transmitting dummy packets if necessary. Therefore, such accuracy can be achieved.

Based on the above two experiments, we can claim that the time slot synchronization accuracy in our system is on the order of 0.4us. 

\subsection{TCP/IP Performance}
\textbf{Method}: In this experiment, we measured the TCP/UDP throughputs and network latency in RPNC. For benchmarking, we compare RPNC with the traditional scheduling (TS) scheme. In TS, the relay schedules nodes A and B to transmit in successive time slots in a non-overlapping manner in the uplink. The PHY-layer parameter configurations (i.e., BPSK and 1/2 coding rate) and the slot duration are the same.

We performed control experiments for SNRs ranging from 6dB to 18dB with 2dB steps. The receive powers of packets from nodes A and B at the relay (i.e., uplink) are adjusted to be balanced, with less than 1dB variation between the two powers. The receive power of the packets from relay at nodes A and B (i.e., downlink) are also adjusted to be balanced. We use the software iPerf \cite{IPERF} to generate saturated traffic to measure the UDP throughputs. The results are averaged over 10 rounds, with each round lasting 300 seconds.

\textbf{UDP Throughputs}: Fig. \ref{fig:UDPThroughput} shows the UDP throughputs of RPNC and TS for different SNRs. We only present the results for the flow from node A to node B. Results for the reverse flow are similar. As can be seen, RPNC outperforms TS by a factor of close to 2 when SNR$\geq$8dB \footnote{When SNR is high, higher-order modulations could be considered for both TS and RPNC. However, our current system is still compute bound, and not communication bound (i.e., use of high-order modulation will not improve the throughput of the current system).Therefore, in this paper we focus on BPSK, and demonstrate the throughput improvement brought about by the reduction of the number of time slots for exchange of two messages between the end nodes from four to two using PNC. The time slot reduction is also valid in PNC systems with higher-order modulations if the systems are communication bounded rather than compute bound. Information theoretically, PNC using lattice codes can achieve rates within 1/2 bit of the cut-set outer bound in two-way relay channel  for all SNR \cite{NamIT2010capacity}. Thus, PNC is not limited to the use of low-order modulations only and it has a fundamental advantage over traditional relay systems.}. 
When SNR$=$6dB, there are still nearly 50\% improvement. The reduced improvement is due to the so-called XOR-CD PNC decoding scheme \cite{PNCSurvey12} at the PHY layer, which is known to have low complexity but poor performance in the low SNR regime. 

When SNR$\geq$14dB, the UDP throughput of RPNC reaches 1.13Mbps, and does not improve further. This is because we use BPSK modulation, and the packet reception rate is nearly 100\% beyond 14dB in our experiment. However, the achieved throughput is close to the optimal PHY-layer throughput (i.e., 1.24Mbps without considering link-layer overhead). We note that the achievable throughput is much smaller than commercial communication systems (e.g., 54Mbps in 802.11a). This is because RPNC only adopts 5MHz bandwidth and BPSK due to computation limitation of USRP/GNU Radio. To further boost throughput, wider bandwidth and higher-order modulations will be considered in the future.

\begin{figure*}[t]
	\begin{minipage}{0.32\linewidth} 
		\centering
		\includegraphics[width=0.99\textwidth]{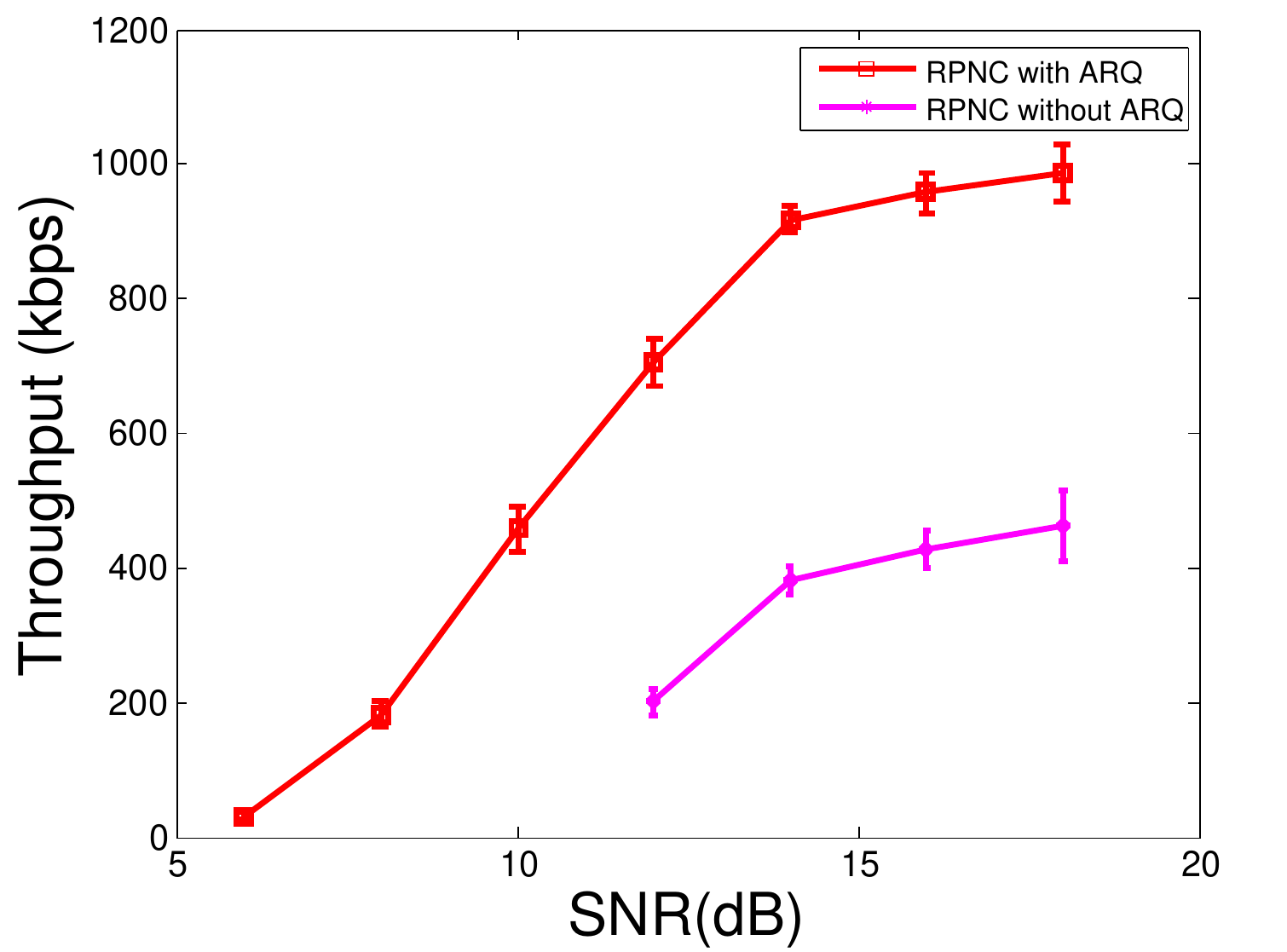}
		\caption{TCP throughputs over different SNRs for RPNC with ARQ and RPNC without ARQ.}\label{fig:TCP-ARQ-Throughput}
	\end{minipage}
	\hfill
	\begin{minipage}{0.32\linewidth} 
		\centering
		\includegraphics[width=0.99\textwidth]{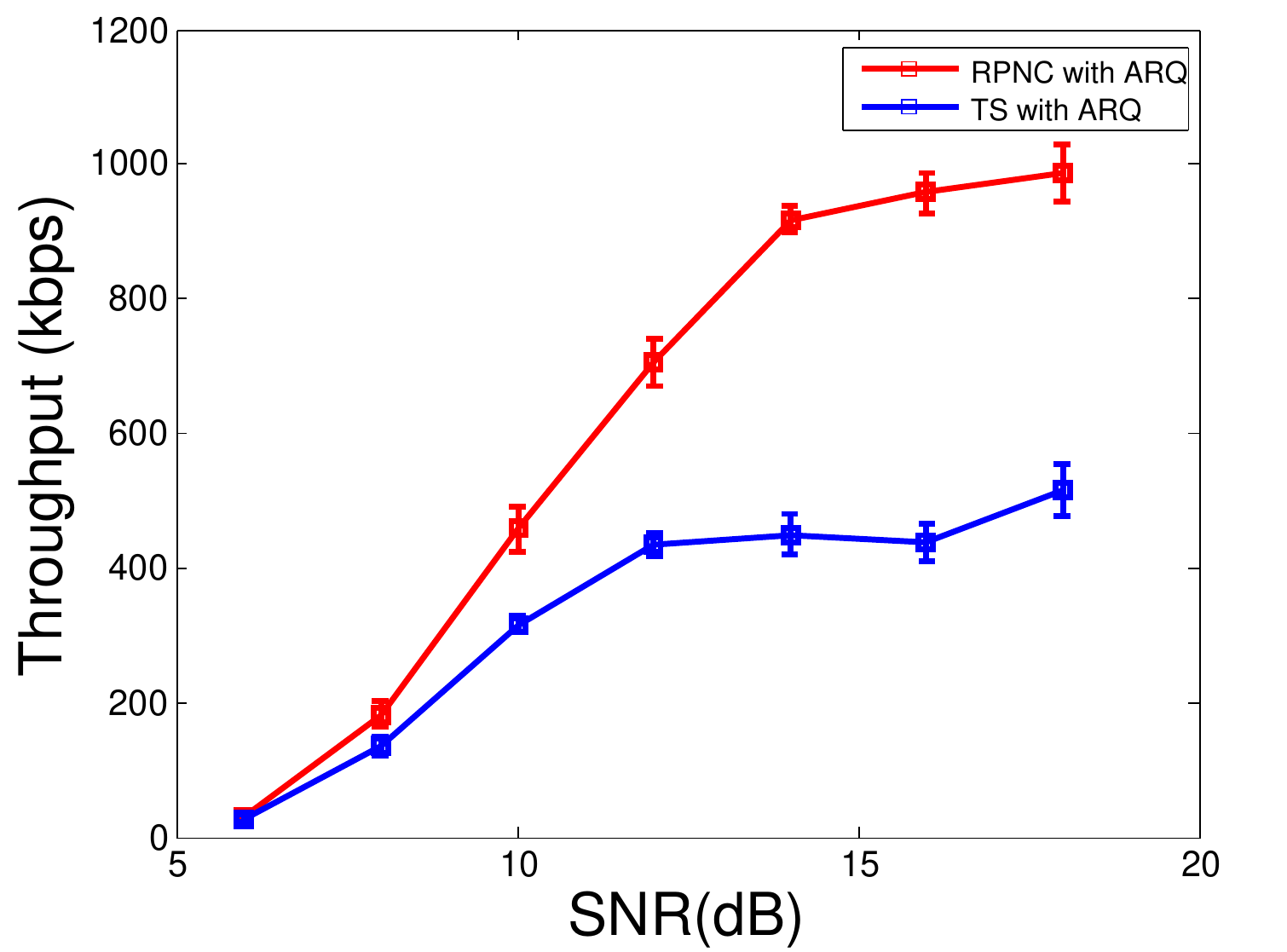}
		\caption{TCP throughputs over different SNRs for RPNC with ARQ and TS with ARQ.}\label{fig:TCPThroughput}
	\end{minipage}
	\hfill
	\begin{minipage}{0.32\linewidth} 
		\centering
		\includegraphics[width=0.99\textwidth]{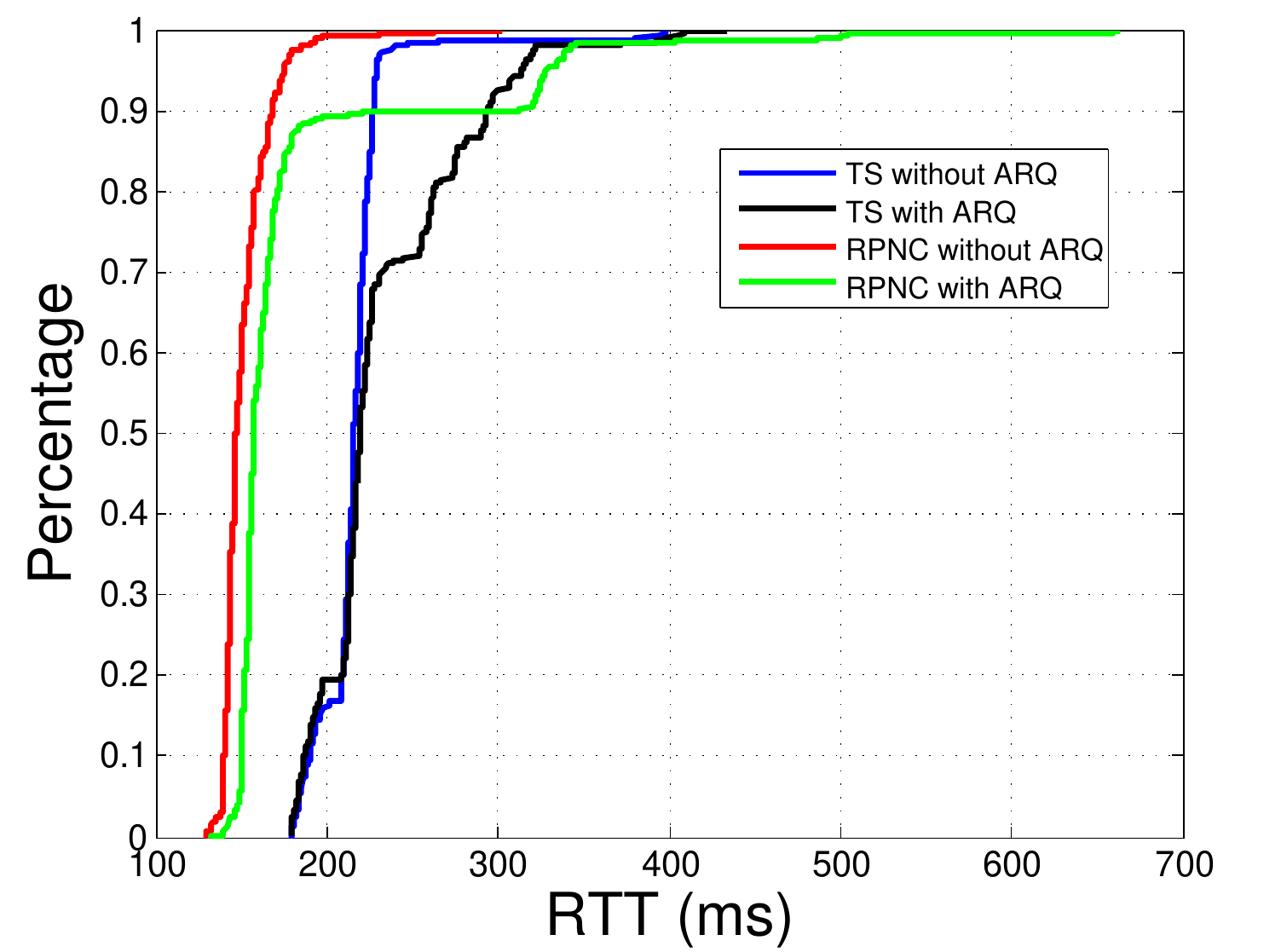}
		\caption{Round trip time for RPNC and TS.}\label{fig:RTT}
	\end{minipage}
\end{figure*}

%
%

\textbf{TCP Throughputs}: Fig. \ref {fig:TCP-ARQ-Throughput} shows the TCP throughputs with link-layer ARQ and without link-layer ARQ for RPNC (See the ARQ protocol in Sec. \ref{sec:llc:arq}). As shown in this figure, for SNR$\geq$14dB, link-layer ARQ improves the TCP performance significantly for all SNRs. For SNR$<$14dB, TCP could not work without link-layer ARQ, because the packet error rate is too high to support TCP \cite{TCPWireless}. However, with link-layer ARQ, TCP can still work in these SNR regimes, although the throughputs degrade.

Fig. \ref{fig:TCPThroughput} shows the TCP throughputs with link-layer ARQ for RPNC and TS. As can be seen, RPNC outperforms TS by a factor of 1.5 to 2.5 when SNR$\geq$ 10dB. When SNR$<$10dB, the improvement is marginal, because the system performance is dominated by the high packet error rate in these SNR regimes, and the reduced time slot in RPNC does not help a lot. 

\textbf{Network Latency}: Fig. \ref{fig:RTT} shows the round trip time (RTT) of RPNC and TS when SNR$=$18dB, with and without link-layer ARQ. We used PING to measure the network latency (over 1000 PING packets). As can be seen, with ARQ, the RTTs increase slightly for both RPNC and TS, because ARQ retransmits packets to recover from packet loss, and thus induces the latency. In addition, RPNC also outperforms TS in terms of network latency, because RPNC allows overlapped transmissions while TS not.

For both RPNC with ARQ and without ARQ, almost all RTTs are less than 200ms. The results show that RPNC is adequate to support real-time applications such as voice communication and video conferencing. 

\section{Related Work} \label{sec:related}
\textbf{Real-time Systems in GPP-based SDR}: Due to the large latency and jitters between USRP and GNU Radio, few high-performance real-time systems are reported on the USRP/GNU Radio platform. Fuxj$\ddot{a}$ger et al. \cite{GR80211Tx} presented a real-time 802.11p OFDM transmitter on USRP/GNU Radio. Bloessl et al. \cite{GR80211Rx,Bloessl14} presented a real-time 802.11a/g/p OFDM receiver on USRP/GNU Radio. Neither is a complete system that supports real TCP/IP applications. To support real applications, our system is designed to ensure link-layer reliability and prevent computation overloading. In addition, we design the time-slotted architecture to address challenges for implementing RPNC. 

Tan et al. \cite{SoraNSDI09} reported an 802.11 a/b/g compliant implementation that supports real TCP/IP applications on the SORA GPP-based SDR platform. Our paper here, on the other hand, focuses on PNC rather than 802.11. Specifically, we design protocols for PNC, and achieve real-time operations in the compute-bound USRP/GNU Radio platform. We believe the demonstrated protocols in RPNC could also benefit the implementation of future compute-bound systems on SORA.

\textbf{PHY-layer Implementation of PNC}: A simplified version of PNC, called analog network coding (ANC) \cite{KattiANC07}, was implemented in TWRN. Recently, Wang and Mao \cite{RANC15} extended ANC to support the QAM modulation, and designed some channel estimation methods for ANC. Although ANC is simple to implement, the disadvantage is that the relay amplifies the noise along with the signal before forwarding the signal, causing error propagation. In this paper, we consider the original PNC based on XOR mapping \cite{PNC06}.  

Lu, Wang, Liew and Zhang \cite{FPNCPhycom12} presented the first PNC implementation based on OFDM, and evaluated the PHY-layer SNR/BER performance through experiments. Wu et al. \cite{DekorsyWSA14} also considered an OFDM-based PNC system, and focused on the problem of carrier frequency offset (CFO) mismatch between the end nodes and the relay. The proposed iterative algorithm was evaluated through FPGA implementation and experiments. 
The works by Chen, Haley and Nguyen \cite{YingAusCTW13} and  Marcum et al. \cite{APNCTWC15} both implemented a single-carrier asynchronous PNC system on USRP, and studied signal processing issues (e.g., CFO estimation and compensation) at the relay.
In contrast,  our paper here focuses on system-level implementation issues. Specifically a  focus of ours is on the design of the MAC layer and the ARQ protocol for TWRN, and we address reliability and real-time challenges for PNC systems on compute-bound GPP-based SDR platform.


\textbf{Link-layer Protocols for PNC-enabled Systems}: Lu et al. \cite{RPNCSRIF13} presented the first real-time implementation of a PNC system, and designed a burst mode MAC protocol where a dedicated beacon frame triggers a burst of uplink transmissions. The current paper extends \cite{RPNCSRIF13} to a more general time-slotted MAC protocol, and demonstrated a complete system supporting real TCP/IP applications.

Mao et al. \cite{RANC-MAC16} proposed a MAC protocol that supports ANC in multihop wireless networks. Wang et al. \cite{SongMACTMC13} proposed a MAC protocol that supports PNC in multihop wireless networks. Nodes in the network collect the queue status of their neighboring nodes through CSMA, and form PNC pairs. Then, the relay node coordinates a PNC pair to start PNC transmissions. He and Liew \cite{ARQTMC15} investigated ARQ designs for PNC systems \cite{PNCBlocksTWC15} in multihop wireless networks, and leveraged  overheard packets together with coded packets to improve throughput.  None of the above work actually proved concepts via a prototype. By contrast, our work here includes a working prototype supporting real TCP/IP applications on a SDR platform.

\textbf{Distributed Time Synchronization}: Network time protocol \cite{NTP} and IEEE 1588 precise time protocol \cite{PTP} are widely used in wired networks to synchronize the clocks of connected nodes. In wireless networks (e.g., Cellular networks), time synchronization is usually implemented through air interface. Unlike Cellular networks, RPNC is based on a distributed wireless architecture (e.g., similar to IEEE 802.11), where there is no control channel for time synchronization. 

IEEE 802.11 \cite{dot11std07} defines the time synchronization function (TSF) to synchronize clocks of all STAs in the same basic service set (BSS). The TSF timer ticks in microsecond ($\mu s$), and the accuracy of the TSF timer is no worse than $\pm$0.01\% \cite{dot11std07}. Given the 100$ms$ beacon interval, the relative clock offset between two STAs in the same BSS is no more than 20$us$. Djukic and Mohapatra \cite{TDMATMC12} improved the synchronization accuracy to 4.21$\mu s \pm$0.02$\mu s$, and designed a TDMA protocol for wireless ad-hoc networks. However, these time synchronization methods are designed for single-user transmission. The accuracy is not adequate for RPNC, since RPNC requires the misalignment of two-user simultaneous transmissions to be within the OFDM CP ($<$3.2$\mu s$). 

Rahul, Hassanieh and Katabi \cite{SourceSync10} achieved 20$ns$ synchronization accuracy, and realized the within CP synchronization requirement for multi-user simultaneous transmissions. However, their implementation is on FPGA, while our implementation is on a GPP-based SDR platform. In particular, we achieve 0.4$\mu s$ synchronization accuracy with pure GPP processing.

\section{Conclusion and Future Work} \label{sec:conclusion}

We have presented and experimentally evaluated the RPNC system that supports real TCP/IP applications over a compute-bound USRP/GNU radio platform. A number of new designs have been put forth and implemented into RPNC to tackle the challenges in realizing a reliable real-time PNC system. Specifically, we designed a distributed time-slotted architecture that achieves $\mu s$-level time synchronization. With the time-slotted architecture, we further designed a new packet identification method (specifically,  a new frame synchronization method that identifies the beginning of a packet) to reduce computation. In addition, we constructed an end-to-end ARQ tailored for RPNC to provide  reliable data delivery in link layer.

\iftrue
The new designs and techniques put forth in this paper are crucial elements that demonstrate the feasibility and usability of PNC in a practical setting. Going forward, a few directions deserve further attention:

\textbf{Higher System Throughput}: Our current system makes use of the BPSK modulation and the 5MHz bandwidth. When SNR is high, higher-order modulations could be considered. In addition, wider bandwidth improves the PHY-layer data rate. The low-complexity PNC decoder for high order QAM and an optimized real-time implementation for wider bandwidth remain to be demonstrated.

\textbf{Link-layer Protocol for Flow Control}: Our current system prefixes the slot duration to avoid overflow at the receiver in accordance with the processing power of our PC. A better way is to fix the slot duration (e.g., to the packet duration), and let the transmitter adaptively adjust the transmission window to match the receiver’s processing capability. Such automatic real-time tuning is more portable across GPPs of different processing capabilities. 

\textbf{Link-by-link ARQ}: Our current prototype implemented an end-to-end ARQ protocol. The ARQ throughput performance could be further improved if we allow the relay to participate in the ARQ process. Joint design of link-by-link ARQ and flow control is also a promising direction for future work.

\fi

\iftrue

\appendices

\section{Frame Synchronization Algorithms for PNC} \label{appendix:sync}

\textbf{Frame Synchronization for Offline PNC}: 
Frame synchronization is an essential function for communication systems. For asynchronous communication systems where a transmitter does not transmit continuously to a receiver over all time, preambles in the packet header facilitate the frame synchronization process. For example, in 802.11 WLAN, a preamble is appended at the beginning of a packet to facilitate frame synchronization. Typically, the preamble contains several repeated short training symbols (STS), and autocorrelation is applied to identify the beginning of a packet.

PNC is a multiuser system. In PNC, the end nodes transmit simultaneously, causing their signals to overlap at the relay. Since there are two end nodes, the relay needs to identify the exact starting samples of the packets of the two end nodes for decoding purposes \cite{FPNCPhycom12}. In addition, the relay needs to determine whether two end nodes transmit, only one end node transmits, or no end node transmits: the appropriate decoder will be invoked depending on the situation. For the offline PNC implementation in \cite{FPNCPhycom12}, cross correlation was adopted. In particular, the two end nodes used identical STS placed at the same position of the frame format. Cross correlation with respect to the unique sequence in STS was applied on all received samples to locate the starting positions. We refer to this processing mechanism as the \emph{exhaustive cross-correlation} algorithm. The description “exhaustive” refers to the fact that the cross correlation algorithm must be run over all samples (the system in \cite{FPNCPhycom12} was not a time-slotted system and thus the beginning of a packet could occur anywhere).

The exhaustive cross-correlation algorithm can be expressed as (here $x^*$ denotes the complex conjugate of $x$)
$$c[n]=\sum_{i=1}^{L} y[n-L+i]x^*[i],$$
where $c[n]$ is the cross correlation result for the $n$-th sample, $y[\cdot]$ are the received samples, and $x[\cdot]$ are the unique sequence in STS of length L. The computation complexity of the cross correlation is $O(NL)$ per time slot \footnote{In this paper, we use a time slot as the basic time unit to compare different synchronization algorithms, since a time slot is a basic element of time-slotted systems. For the unslotted system in \cite{FPNCPhycom12}, it is to be understood that “time slot” is just a time unit for normalization purposes and not that the system is a time-slotted system.}, where $N$ is the number of samples per time slot. In \cite{FPNCPhycom12}, we computed the cross correlation offline. Cross correlation is a burden for a real-time system because of its high complexity. 

\begin{figure}
	\centering
	\includegraphics[width=0.5\textwidth]{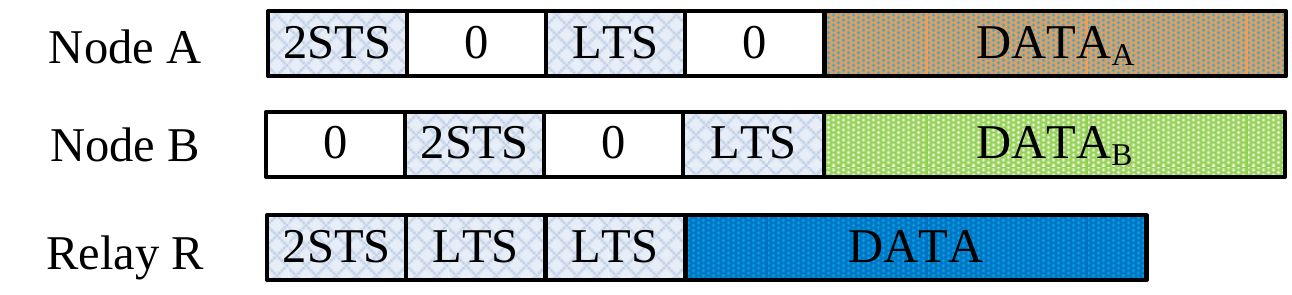}
	\caption{Frame formats for the end nodes and the relay in the time domain: the beginnings are preambles, including short training symbols (STS) and long training symbols (LTS).}\label{fig:FrameFormat}
\end{figure}

\textbf{A Straightforward Modification to Enable Real-time PNC}: 
For real-time PNC systems, a straightforward modification to \cite{FPNCPhycom12} is for the two end nodes to use non-overlapping STS, and to use autocorrelation to identity a rough beginning of a packet. After a rough beginning is identified, then cross-correlation can be used to locate a precise beginning in the neighborhood of the rough beginning. This reduces the amount of cross correlation to be performed, hence the complexity is also reduced since cross correlation is more computationally intensive than autocorrelation.  We call this modification the \emph{exhausive auto-correlation narrow cross-correlation} algorithm. Since it is a common method in real systems \cite{Sams01}, henceforth we will simply refer to it as the \emph{standard cross-correlation} algorithm.

Fig. \ref{fig:FrameFormat} shows the frame formats for the end nodes (i.e., the uplink frame format) and the relay (i.e., the downlink frame format) that support the standard cross-correlation algorithm. The PHY layer of RPNC is based on OFDM, with parameters (e.g., the subcarrier map) similar to those in 802.11. The PHY-layer frame formats of the two end nodes are modified so that they use non-overlapping time-domain STS for multi-user identification, use non-overlapping long training symbols (LTS) for multi-user channel estimation, and transmit the data payloads as overlapped OFDM symbols. There are two repeated STS for the 2STS field shown in Fig. \ref{fig:FrameFormat} to facilitate the autocorrelation computation. Each of the two repeated STS has length $L$. In addition to the two STS of length $2L$, there is a CP of length $C$ in the 2STS field. Overall, the 2STS field lasts for one OFDM symbol duration. The relay uses preambles with one fewer OFDM symbol  duration than the end nodes, omitting one of the 2STS fields, since the relay is the only transmitter in the downlink.

The frame synchronization algorithm is modified as follows. The relay first computes the autocorrelation 
$$a[n]=\sum_{i=1}^{L} y[n-2L+i]y^*[n-L+i]$$
to identify a rough beginning of a preamble, and then performs  the cross correlation of the received signals with respect to STS, 
$$c[n]=\sum_{i=1}^L y[n-L+i]x^*[i],$$
to locate the exact position in the neighborhood of the rough beginning. The cross correlation computation will not be triggered unless it is alerted by the autocorrelation computation first.

This method has a lower implementation complexity than the exhaustive cross-correlation method. The computation complexity of autocorrelation is $O(N)$ per time slot (exploiting the fact $a[n]=a[n-1]+y[n-L]y^*[n]-y[n-2L]y^*[n-L]$). The computation complexity of cross correlation is $O(CL)$ per packet, since the cross correlation is applied on signal samples of length $C$ in the vicinity of a preamble detected by autocorrelation, where $C$ is the length of CP. Typically, $N \gg CL$. Therefore, the complexity of this method is dominated by $O(N)$. 


\begin{figure}
	\centering
	\includegraphics[width=0.5\textwidth]{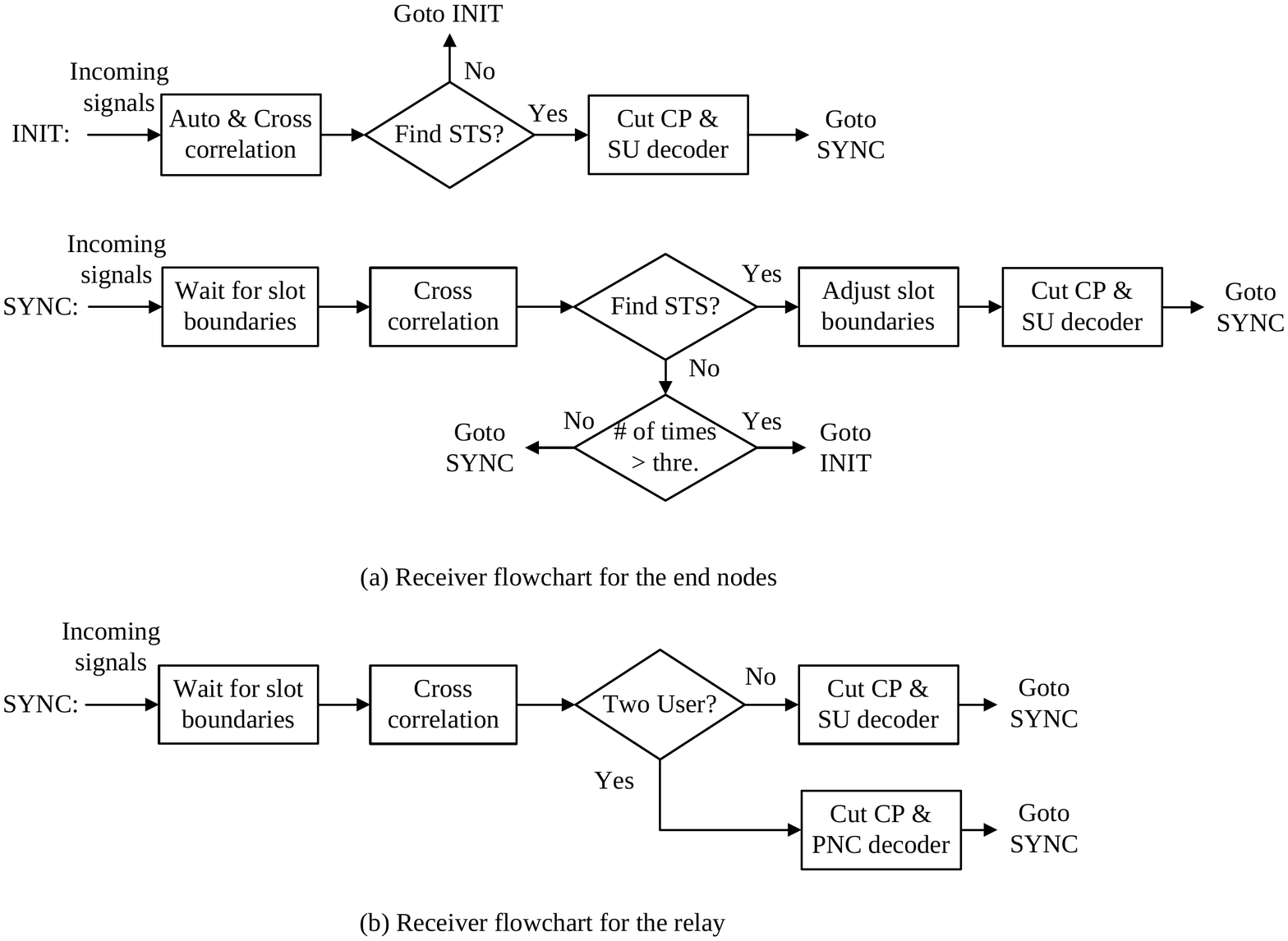}
	\caption{The receiver flowcharts of the end nodes and the relay at the PHY layer.}  \label{fig:PHY-Flowchart}
\end{figure}
	
\textbf{Frame Synchronization for Time-slotted RPNC}:
Based on the design principles discussed in Sec. \ref{sec:phy:framesync}, we next present the receiver flowcharts for the narrow cross-correlation algorithm at the end nodes and at the relay in Fig. \ref{fig:PHY-Flowchart}. There are two states in the flowcharts: \emph{INIT} and \emph{SYNC}. In the \emph{INIT} state, the receiver has not acquired the slot boundaries yet; in the \emph{SYNC} state, the receiver has acquired the slot boundaries. Recall that the relay does not need slot synchronization, and therefore only the \emph{SYNC} state exists for the relay. 
	
For the end nodes, depending on which state the receiver is in, the incoming signals pass through different paths. In the \emph{INIT} state, the receiver first executes the standard cross-correlation algorithm, and cuts CP for decoding if a preamble is found. In the \emph{SYNC} state, the receiver performs cross correlation in the vicinity of the next slot boundary; if an STS is found, the receiver adjusts slot boundaries by Eq. (\ref{eqn:a1}) and (\ref{eqn:b}), and cuts CP for decoding; otherwise, the receiver switches back to the \emph{INIT} state if the number of successive time slots in which an STS is not found exceeds a certain threshold (see Sec. \ref{sec:phy:framesync}). 
	
For the relay, the receiver performs cross correlation at the next slot boundary to check whether each end node transmits a packet. If so, it cuts CP accordingly and invokes the appropriate OFDM decoder (i.e., single-user or PNC).
	
In the above receiver flowcharts, the main computation lies in the cross correlation performed at slot boundaries (if the boot-up cost of the auto-and-cross algorithm in the \emph{INIT} state can be ignored). Therefore, the overall complexity of the synchronization at the PHY layer is $O(CL)$ per time slot: this is a significant improvement over the straightforward modification (i.e., the standard cross-correlation algorithm) with $O(N)$ computation complexity and the previous PNC prototype (i.e., the exhaustive cross-correlation algorithm) with $O(NL)$ computation complexity \cite{FPNCPhycom12}.

\section{Error Detection for Relay} \label{appendix:crc}


We modify the CRC-32 mechanism defined in IEEE 802.11 \cite{dot11std07} to do error detection at relay. To begin with, we introduce the conventional CRC-32 mechanism in 802.11.

Let $\tilde{P}(x)$ be the polynomial representation of the packet payload with length $k$ bits.It is a polynomial of degree $k-1$. The 802.11 standard defines $L(x)=x^k \times (x^{31}+x^{30}+x^{29}+\cdots+x+1)$, and the generator polynomial $G(x)$ of degree 32.
The CRC field of the packet is computed as 
	\begin{equation} \label{eqn:crc-i}
	CRC(x)=\overline{[L(x) \oplus P(x)] \bmod G(x)},
	\end{equation}
where $P(x)=x^{32}\tilde{P}(x)$. $CRC(x)$ is a polynomial of degree 31.

The sender then packs the bits in the two polynomials $\tilde{P}(x)$ and $CRC(x)$ into a packet $[\tilde{P}(x)~CRC(x)]$, and transmits them to the receiver. Assuming the receiver receives and decodes the bits as $[\tilde{P}_{rx}(x)~CRC_{rx}(x)]$, it then computes the expected CRC field as in Eq. (\ref{eqn:crc-i}),
	\begin{equation} \label{eqn:crc-ii}
	CRC_{cal}(x)=\overline{[L(x) \oplus P_{rx}(x)] \bmod G(x)},
	\end{equation}
where $P_{rx}(x)=x^{32} \tilde{P}_{rx}(x)$, and compares $CRC_{rx}(x)$ and $CRC_{cal}(x)$: if they are the same, the receiver declares a correct packet has been received; otherwise, the receiver declares an erroneous packet has been received. 

We now explain how we modify the conventional CRC mechanism above for the uplink of RPNC. Let $[\tilde{P}^A(x)~CRC^A(x)]$ and $[\tilde{P}^B(x)~CRC^B(x)]$ be the packets (of the same length) transmitted by nodes A and B respectively, and $[\tilde{P}^{A \oplus B}_{rx}(x)~CRC^{A \oplus B}_{rx}(x)]$ be the packet decoded by the relay. For XOR decoding, we can treat nodes A and B as a virtual node transmitting $[\tilde{P}^{A \oplus B}(x)~CRC^{A \oplus B}(x)]$, where $\tilde{P}^{A \oplus B}(x)=\tilde{P}^A(x) \oplus \tilde{P}^B(x)$ and $CRC^{A \oplus B}(x) = CRC^A(x) \oplus CRC^B(x)$. 

We modify step in Eq. (\ref{eqn:crc-ii}), and compute $CRC^{A \oplus B}_{cal}(x)$ in the following manner:
\begin{equation} \label{eqn:crc-iii}
CRC^{A \oplus B}_{cal}(x)=P^{A \oplus B}_{rx}(x) \bmod G(x),
\end{equation}
where $P^{A \oplus B}_{rx}(x) = x^{32} \tilde{P}^{A \oplus B}_{rx}(x)$. 

By comparing $CRC^{A \oplus B}_{rx}(x)$ and $CRC^{A \oplus B}_{cal}(x)$, the relay could then perform error detection on the XOR packet as in the conventional CRC. Note that, at the transmitters’ sides, nodes A and B still perform Eq. (\ref{eqn:crc-i}) with the inclusion of  in the CRC computation for the packets that they transmit. However, the subtlety here is that, unlike a standard 802.11 receiver, the relay must exclude $L(x)$ and the complementary operation in Eq. (\ref{eqn:crc-iii}) in the CRC computation of the XOR packet. The reason for doing so is as follows. Since both nodes A and B include $L(x)$ and the complementary operation in their CRC computation, this is equivalent to a virtual node sending the XOR CRC given by
\begin{align*}
CRC(x) &=\{\overline{[L(x) \oplus P^A(x)] \bmod G(x)}\} \\
&~~~~~~~~~~~~~~~~~~ \oplus \{\overline{[L(x) \oplus P^B(x)] \bmod G(x)}\} \\
&=\{[L(x) \oplus P^A(x)] \bmod G(x)\} \\
&~~~~~~~~~~~~~~~~~~ \oplus \{[L(x) \oplus P^B(x)] \bmod G(x)\} \\
&=[L(x) \oplus P^A(x) \oplus L(x) \oplus P^B(x)] \bmod G(x) \\
&=[P^A(x) \oplus P^B(x)] \bmod G(x).
\end{align*}
In other words, the two $L(x)$ and complementary operations nullify each other through the XOR operation. 

We note that this mechanism only applies if XOR packets are received. If only one user transmits in the uplink, the relay reverts back to the conventional 802.11 CRC mechanism for error detection.

%

%

\ifCLASSOPTIONcompsoc
  \section*{Acknowledgments}
\else
  \section*{Acknowledgment}
\fi


This work is supported by AoE grant E-02/08 and the General Research Funds Project Number 14204714, established under the University Grant Committee of the Hong Kong Special Administrative Region, China. This work is also supported by the China NSFC grants (Project No. 61271277 and No. 61501390).

\fi




%
%
%
\bibliographystyle{IEEEtran}
\bibliography{pnc}

%

%
%
%




\end{document}